\documentstyle[aps,pre,epsf,floats]{revtex}
\def\FigureInText#1{#1}
\def\TextEndsHere{\end{document}}
\begin{document}
\twocolumn[\hsize\textwidth\columnwidth\hsize\csname@twocolumnfalse%
\endcsname


\draft
\title  {Flowing sand - a possible physical realization of 
	 Directed Percolation}

\author {Haye Hinrichsen$^1$, Andrea Jim\'enez-Dalmaroni$^{2,3}$, Yadin Rozov$^2$, 
          and Eytan Domany$^2$\\[2mm]}

\address{$^{1}$
        Max-Planck-Institut f\"ur Physik komplexer Systeme,
        N\"othnitzer Stra\ss e 38, 01187 Dresden, Germany \\       
 	$^{2}$ 
	Department of Physics of Complex Systems,
        Weizmann Institute of Science, Rehovot 76100, Israel\\
	$^{3}$ 
	University of Oxford, Department of Physics - 
	Theoretical Physics, 1 Keble Road, Oxford OX1 3NP, U.K.       
 }
\date   {September 27, 1999}
\maketitle

\begin{abstract}
A simple model for flowing sand on an inclined plane is
introduced. The model is related
to recent experiments by Douady and Daerr 
{[}Nature {\bf 399}, 241 (1999){]} and reproduces some of the 
experimentally observed features. Avalanches of intermediate
size appear to be compact, placing the critical behavior of the
model into the universality class of compact directed percolation.
On very large scales, however, the avalanches break up
into several branches leading to a crossover
from compact to ordinary directed percolation.
Thus, systems of flowing granular matter on an inclined
plane could serve as a first physical realization 
of directed percolation.\\
\end{abstract}

]

%

\renewcommand{\thefootnote}{\arabic{footnote}}

\section{Introduction}

Directed Percolation (DP) is perhaps 
the simplest model that exhibits a
non-equilibrium phase transition between an 
``active'' or ``wet'' phase and an
inactive ``dry'' one~\cite{DP}. 
In the latter phase the system is 
in a single ``absorbing'' state; once it reaches the
completely dry state, it will always stay there. 

Interest in DP mainly stems from  
{\it universality} of the associated
critical behavior. It is believed 
that transitions in all models
with an absorbing state belong to the 
DP universality class (unless there are some
special underlying symmetries). 
DP exponents were measured for 
an extremely wide variety of models. Even though
the exponents have not yet been 
calculated analytically, their values
(especially in 1+1 dimensions) are
known with very high precision~\cite{Jensen}.

Despite the preponderance of models 
in the DP universality class, so far 
no physical system has been found 
to exhibit DP behavior. Indeed, as noted
by Grassberger, 
\begin{quote}
``{\it ...there is 
still no experiment where the critical 
behavior of DP
was seen. This is a very strange 
situation in view of the vast and
successive theoretical efforts 
made to understand it. Designing and
performing such an experiment 
has thus top priority in my list of open
problems''}~\cite{Grassberger96}.
\end{quote} 
The purpose of this paper is to point out that a  
simple system of sand flow on an 
inclined plane, that  has recently been
introduced and studied by Daerr and  
Douady (DD), may well be the first 
physical realization of a transition in the DP universality 
class~\cite{DouadyDaerr98,DaerrDouady99}. 
In Sec.~II we describe these experiments in fair detail. 
The data presented by DD is of qualitative value
and raises serious questions regarding
the applicability of DP. In particular, 
the observed shapes of wet 
clusters differ from those seen in 
standard DP simulations; they are much more
compact. Since the corresponding model, 
called Compact Directed Percolation
(CDP), is unstable against perturbations 
towards the standard DP behavior
\cite{DomanyKinzel},
the latter is the generic case 
expected to occur (if no parameters were
fine-tuned to place the system in the CDP class).  

This motivated us to look for a   
simple model which is defined in terms of 
dynamic rules that can plausibly be 
related to the experiments and, at the same time, 
exhibit features that look like the   
experimentally observed ones. 
Whether the transition exhibited by such a model 
does belong to the DP universality 
class remains to be investigated.

Such a model is introduced in Sec.~III.
It is a {\it directed sandpile} 
model, which is simpler than the one introduced
and analyzed by Tadic and Dhar~\cite{TadicDhar97}; 
here the system is reset to a uniform initial state
after each avalanche.  
In Sec.~IV we show the outcome of some
simulations. The avalanches (observed in the active phase)  
reproduce the experimental observations quite well. 
We establish the existence
of a transition from an active to an inactive phase.  
However, the critical behavior 
extracted from these figures does not
seem to be in the DP universality 
class, rather, it seems close to
CDP.  As it turns out, this CDP type 
critical behavior is only a transient:
the true critical
behavior {\it is} of the DP type, 
but can only be seen after a very long
crossover regime, in which 
the exponents are those of CDP. 
This observation is based on 
a careful numerical study, 
which is presented in Secs.~V and VI. 

Our conclusion is that the DD 
experiment does serve as a possible
realization of a DP-type transition. 
Observation of DP exponents may be tricky as
a substantial crossover regime may mask the true 
critical behavior, and one should try to find methods
to shorten this regime.

Finally we should note that the DD 
system is a simple case of Self Organized
Criticality (SOC). Without any 
fine tuning, the system ``prepares itself'' 
at the critical point of a DP type 
transition. The way in which this happens
differs from standard
SOC models~\cite{Bak} in which a slow 
driving force (acting on a much time 
scale smaller than that of the system's 
dynamic response) causes evolution to 
a critical state. In the present case 
avalanches are started by hand one by one. 

\section{The Douady-Daerr Experiment}

The experimental apparatus consists 
of an inclined plane (size of about
$1m$) covered by a rough velvet cloth; 
the angle of inclination  
$\varphi _{0}$ can
be varied. Glass beads (e.g.``sand'') 
of diameter $250$-$425 \ \mu$m~\cite{DouadyDaerr98} 
are poured uniformly at the
top of the plane and flow down while
a thin layer of thickness   
$h=h_{d}(\varphi _{0})$, consisting 
of several monolayers, settles and
remains immobile. At this thickness the sand 
is {\it dynamically stable}; the value 
of $h_d$ decreases with an increasing 
angle of inclination. 

\FigureInText{
\begin{figure}
\epsfxsize=85mm
\centerline{\epsffile{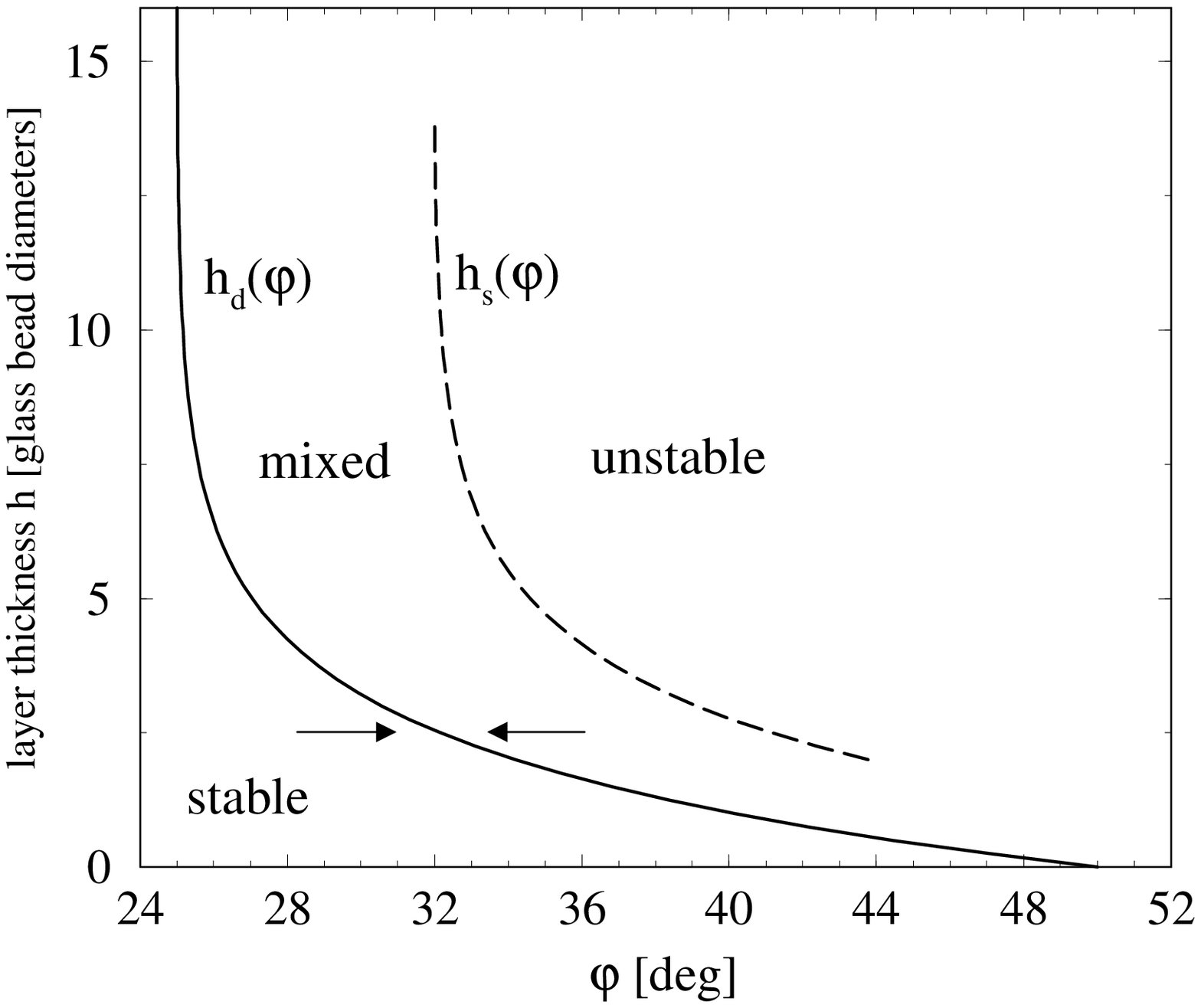}}
\caption{
\label{FigExperiment}
Schematic stability diagram of the DD-experiment.
A layer of thickness $h$ is dynamically stable
below a certain threshold $h_d(\varphi)$ (solid
line). Due to friction forces non-moving layers remain
stable in the mixed region below static stability 
limit $h_s(\varphi)$ (dashed line).
In the present work we investigate the properties
in the vicinity of the dynamic phase transition
line, as indicated by the arrows.
}
\end{figure}
}

For each $\varphi_0$ there exists another thickness $h_s$ 
with $h_s(\varphi _{0}) > h_{d}(\varphi _{0})$, 
beyond which a {\it static} layer
becomes unstable. Hence there exists a region 
(see Fig.~\ref{FigExperiment}) in the 
$(\varphi,h)$ plane, in which a static 
layer is stable but a flowing one is
unstable. We can now take the system, 
that settled at $h_{d}(\varphi _{0})$, 
and increase its angle of inclination
to $\varphi$, staying within this 
region of mixed stability. The layer will 
not flow spontaneously, but if we 
disturb it at the top, generating
a flow near the perturbation, the flow
will persist and an avalanche will be generated, leaving 
behind a layer of thickness
$h_d(\varphi)$. These avalanches 
had the shape of a fairly regular triangle,
with opening angle~$\theta$.
As the increment of the inclination 
\[
\Delta \varphi = \varphi -   \varphi_0
\]
decreases, the value of $\theta(\Delta \varphi)$ 
decreases as well and the area
affected by the avalanche decreases, vanishing as 
$\Delta \varphi \rightarrow 0$. 
This calls for testing a power law behavior
of the form
\begin{equation}
\theta \sim (\Delta \varphi)^x \,.
\label{eq:powerlaw}
\end{equation}
If instead of increasing $\varphi$ 
we lower the plane, i.e.,  go to
$\Delta \varphi <0$, our system, 
whose thickness is $h_{d}(\varphi _{0})$,
is  below the present thickness of 
dynamic stability, $h_{d}(\varphi )$.
We believe that in this case
an initial perturbation will not 
propagate, it will rather die out 
after a certain time (or beyond a certain size 
$\xi_\parallel$ of the transient avalanche). As the deviation 
$ \vert \Delta \varphi \vert$ decreases, 
we expect the size of the transient active region to increase, i.e., 
the decay length should grow according to a power law
\begin{equation}
\xi_\parallel \sim (-\Delta \varphi)^{-\nu_\parallel} \,.
\label{eq:powerlaw2}
\end{equation}
Hence, by pouring sand at inclination 
$\varphi_0$, DD  produced a {\it
self-organized critical system}. The 
system is precisely at the borderline
(with respect to changing the angle) 
between a stable regime $\varphi < \varphi_0$ 
in which perturbations die out and an unstable one, 
$\varphi > \varphi_0$, where perturbations persist and spread.  

Once this connection has been made, 
it is natural to associate this system with
the problem of DP. Denote by~$p$ 
either the site or bond percolation probability
and by $p_c$ its critical value 
(i.e., for $p > p_c$ the system is in the active phase). 
We associate the change in tilt with $p - p_c$, 
assuming that near the angle of preparation the behavior
of the sand system is related to a DP problem with
\begin{equation}
\Delta \varphi = \varphi - \varphi_0 \propto p - p_c \,.
\label{eq:p_pc}
\end{equation}
Hence, the exponent $\nu_\parallel$ should be 
compared with the known values for
DP and CDP. The exponent $x$ in Eq.~(\ref{eq:powerlaw}) 
can also be measured and compared with 
\begin{equation}
\tan \theta \sim \xi_\perp / \xi_\parallel \sim 
(\Delta \varphi )^{\nu_\parallel - \nu_\perp } \,.
\label{eq:Dnu}
\end{equation}

\section{The Model}

Our aim is to write down a simple model based on the physics of 
flowing sand. We adopt the observation made by DD, that in 
the regime of interest (i.e., for tilt angles close to 
$\varphi_0$) grains of the top layer of sand 
rest on grains of the layer below (rather 
than on other grains of the top layer)\footnote{
This holds for $\varphi < \varphi_0$ and also for  
$\varphi > \varphi_0$, as long as we 
stay within the region of mixed stability.}. 
Hence the lower layers provide for
the top one a kind of {\it washboard potential}, 
as depicted in Fig.~\ref{FigPotential}.

\FigureInText{
\begin{figure}
\epsfxsize=70mm
\centerline{\epsffile{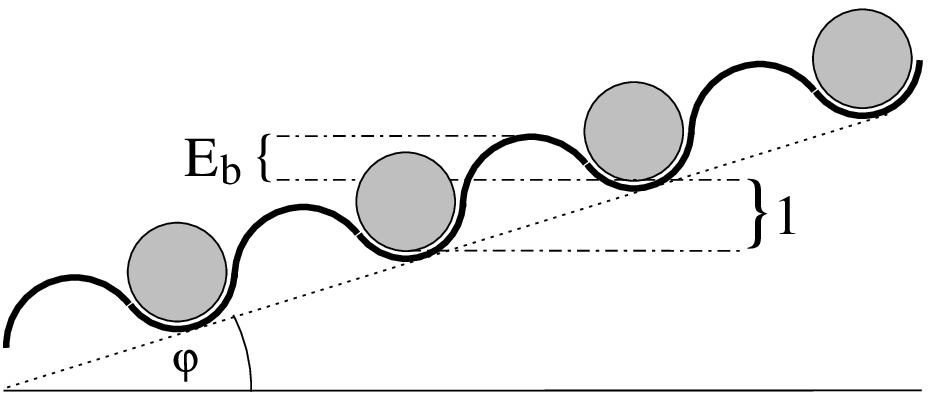}}
\vspace{2mm}
\caption{
\label{FigPotential}
The top layer of sand in an effective washboard potential.
}
\end{figure}
}

We further assume that only the top layer participates in 
an avalanche and therefore place the grains of this 
layer on the sites of a regular square 
lattice\footnote{We chose to work with a 
square lattice, but could have used a triangular one as well, 
with each site communicating with two neighbors above and two below.}
(see Fig.~\ref{FigSquareLattice}). 
At any given time a particular horizontal row  
of grains may become active, while at the next time 
step the activity may be transferred to the row beneath.
The physical picture that underlies the model is as follows. 
A grain $G$ may become active if at least one of the neighboring 
grains in the row above it has been active at the previous 
time step. These grains may then transfer energy to
$G$; if $\Delta E(G)$, the total energy transferred to $G$,
exceeds the barrier $E_b$ of the washboard, 
$G$ becomes active. An active grain ``rolls down'' at the next 
time step and collides with the grains of the next row. 
The energy it brings to  these collisions
is $1~+~\Delta E(G)$, where 1 is the 
potential energy due to the height 
difference between two consecutive rows.
A fraction $f$ of its total energy is
dissipated, while the rest is divided 
stochastically among its three neighbors from
the lower row. 

The model is hence defined in terms of two variables; an activation variable,
\[
S^t_i = \left\{ 
\begin{array}{ll}
1 \qquad  &  \mbox{if grain $(t,i)$ active,} \\
0         &  \mbox{otherwise,}
\end{array}
\right.
\] 
and an energy variable $E^t_i$. The index $t$ denotes rows of our
square lattice {\it and } time; at time $t$ we update the states of the grains
belonging to row $t$. Energy is measured in units of the  difference
between two successive minima of the potential 
(see Fig.~\ref{FigPotential}). The model is controlled by two
parameters, namely
\begin{tabbing}
\= $\qquad E_b\,, \qquad$   \= the barrier height, and\\
\> $\qquad f\,,$     \> the fraction of dissipated energy. 
\end{tabbing}
The dynamic rules of our model are defined in terms of these variables 
and parameters as follows. For given  values of activities $S_i^{t}$ and 
energies $E_i^{t}$  we first calculate the energy transferred 
to the grains of the next row $t+1$.
To this end we generate for each active site $S_i^t=1$ three random
numbers, $z_i^t(\delta )$ (with $\delta = \pm1,0$) in a way that
\begin{equation}
\sum_{\delta = \pm 1,0} z_i^t(\delta ) = 1 \,.
\label{eq:xi}
\end{equation}
The energy transferred to grain $(t+1,i)$ is then given by
\begin{equation}
\Delta E_i^{t+1}~~=~~(1-f)\sum_{\delta=\pm 1,0} S^{t}_{i-\delta} 
~ E^{t}_{i-\delta}
~ z_{i-\delta}^{t}(\delta) \,.
\label{eq:DEt}
\end{equation}
\FigureInText{
\begin{figure}
\epsfxsize=70mm
\centerline{\epsffile{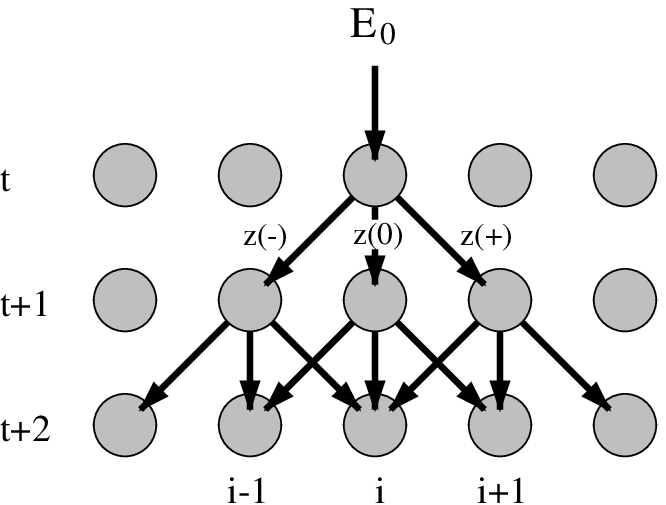}}
\vspace{1mm}
\caption{
\label{FigSquareLattice}
Energy transfer between grains on a square lattice.
}
\end{figure}
}

\noindent
The values of these energies determine the  activation 
of the grains of row $t+1$:
\begin{equation}
S^{t+1}_i= \left\{
\begin{array}{cc}
1   &   \mbox{if $~~~\Delta E^{t+1}_i > E_b$} \,, \\
0   &   \mbox{if $~~~\Delta E^{t+1}_i \leq E_b$} \,.
\end{array}
\right.
\label{eq:St}
\end{equation}
The energies of the active grains are set according to
\begin{equation}
E_i^{t+1}~ = ~ S_i^{t+1} ~ (1 + \Delta E_i^{t+1}) \,.
\label{eq:Et}
\end{equation}
The meaning of these rules, in words, is obvious:
the energy of site $i$ at time $t+1$ 
is obtained by identifying, among its
three neighbors of the preceding row, 
those sites (or grains) that were active at time $t$.
At each such active site $(t,i)$ we generated three 
random numbers $z_i^{t}(\delta)$ which represent 
the fraction of energy transferred  from  the grain 
at site $(t,i)$ to the one at $(t+1,i+\delta)$. 
We add up the energy contributions from these 
active sites; the fraction $1-f$ is {\it not}
dissipated and compared to the barrier height $E_b$. 
If the acquired energy $\Delta E_i^{t+1}$ exceeds $E_b$ , 
site $(t+1,i)$ becomes active, rolls over the barrier
bringing to the collisions (at time $t+2$) the acquired energy 
calculated above {\it and} its excess potential energy (of value 1). 

\section{Short-time simulations  
and qualitative discussion of the transition}

%
%
\FigureInText{
\begin{figure}
\epsfxsize=75mm
\centerline{\epsffile{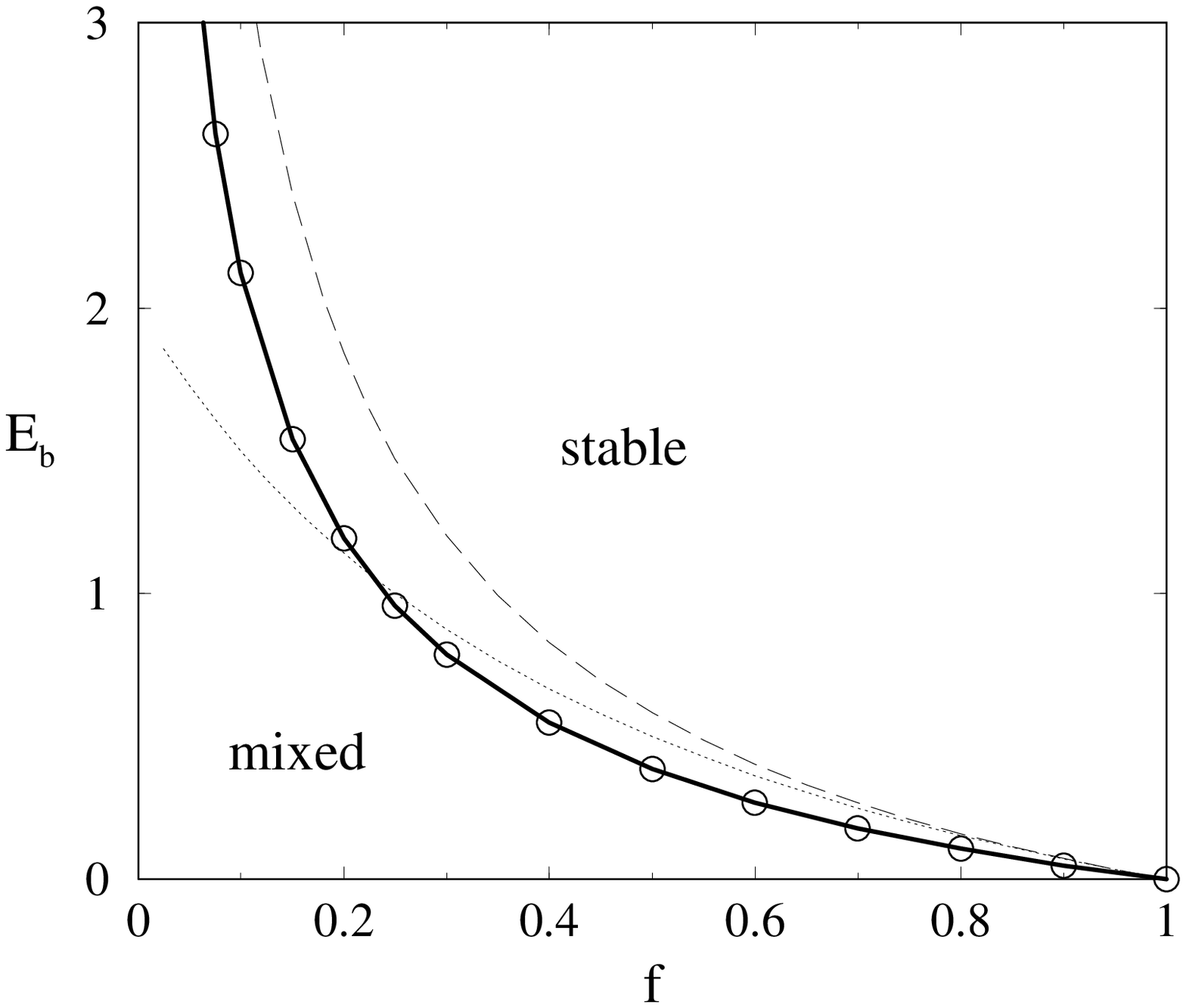}}
\caption{
\label{FigPhaseTrans}
Phase diagram of the model for flowing sand. 
The full line represents the phase transition 
line. The mean-field approximations
of Eqs.~(\ref{SimpleMeanField}) 
and~(\ref{ImprovedMeanField}) are
shown as dotted and dashed lines, 
respectively. 
}
\end{figure}
}

Let us consider the behavior of our model as we vary $E_b$ at a 
fixed value of the dissipation. We expect that for small values of 
$E_b$ an active grain will activate the grains below with high probability;
avalanches will propagate downhill and also spread sideways. For a 
strongly localized initial activation we should, therefore, observe 
activated regions of triangular shape. As $E_b$ increases,
the rate of activation decreases and the opening angle 
$\theta$ of these triangles should decrease, 
until $E_b$ reaches a critical value $E_b^c$, beyond which initial
activations die out in a finite number of time steps (or rows).
These expectations are indeed borne out by simulations of the model:  
the critical value $E_b^c$  depends on the dissipation $f$ and the resulting
phase transition line is shown in Fig.~\ref{FigPhaseTrans}
as a solid line.

In order to understand this transition qualitatively, let us 
consider a simple mean-field type approximation, 
in which all stochastic
variables are replaced by their average values.
 
We consider an edge separating an active
region from an inactive one at time $t$: sites
to the left of $i$ and $i$ itself are wet, 
whereas $i+1,i+2,...$ are dry.  
Will the rightmost wet site be wet or dry 
at the next time step? Assuming that all wet 
sites at time $t$ have the same energy $E^t$,
in our mean-field type estimate the energy 
delivered to site $i$ at time $t+1$ is
\begin{equation}
\Delta E^{t+1}_i=\frac{2}{3}(1-f)(1+\Delta E^{t}) \,,
\label{eq:MFE}
\end{equation}
where we set in Eq.~(\ref{eq:DEt})
all $z(\delta)=1/3$. At the critical point we expect all energies 
just to be sufficient to go over the barrier; hence set  
$\Delta E_i^{t+1}=\Delta E^t=E_b^c$  
in Eq.~(\ref{eq:MFE}). Solving the resulting equation yields
\begin{equation}
E_b^c=\frac{2(1-f)}{1+2f} \,.
\label{SimpleMeanField}
\end{equation}
In Fig.~\ref{FigPhaseTrans} this rough estimate of the 
transition line is shown as a dotted line. 

This simple calculation captures the physics of the problem.
However, it is easy to improve it in the 
following way. As before, we assume the 
energy of toppling grains to be distributed equally
among the three neighbors of the subsequent row.
However, we no longer assume all active sites to
carry the {\em same} energy, instead we compute the energy
profile at the edge of a cluster. To this end let 
us consider a semi-infinite cluster with 
$S_i^t=1$ for $i\leq 0$ and $S_i^t=0$
for $i>0$. According to Eq.~(\ref{eq:DEt}), we are 
looking for a stationary solution of the equation 
of motion
$$
\Delta E_i^{t+1}=\frac{1-f}{3} \left\{
\begin{array}{ll}
3+\Delta E_{i-1}^t+\Delta E_{i}^t+\Delta E_{i+1}^t \, & \mbox{if } i<0 \\
2+\Delta E_{-1}^t+\Delta E_{0}^t \, & \mbox{if } i=0 \\
0 \qquad & \mbox{if } i>0 
\end{array}
\right.
\nonumber
$$
where $\Delta E_0^t=E_b^c$. The corresponding stationary
solution reads
\begin{equation}
\Delta E_i^{stat}=E_{bulk}-E_{gap} \,\exp{(ai)}\,, \qquad (i\leq 0)
\end{equation}
where 
\begin{eqnarray}
E_{bulk}&=&(1-f)/f\,, \nonumber \\
E_{gap}&=&\frac{2+f-\sqrt{12f-3f^2}}{2f(1-f)}\,, \\
a&=&\mbox{arccosh} \frac{2+f}{2-2f} \nonumber \,.
\end{eqnarray}
Thus, the critical threshold is given by the expression
\begin{equation}
E_b^c=\frac{2f^2-5f+\sqrt{12f-3f^2}}{2f(f-1)}
\label{ImprovedMeanField}
\end{equation}
which slightly improves the mean field 
result~(\ref{SimpleMeanField}), especially for small values of
$f$ (see dashed line in Fig.~\ref{FigPhaseTrans}).The energy 
profile decreases at the edges of the cluster
and saturates in the bulk at $E_{bulk}$,
as shown in Fig.~\ref{FigProfile}.

\FigureInText{
\begin{figure}
\epsfxsize=75mm
\centerline{\epsffile{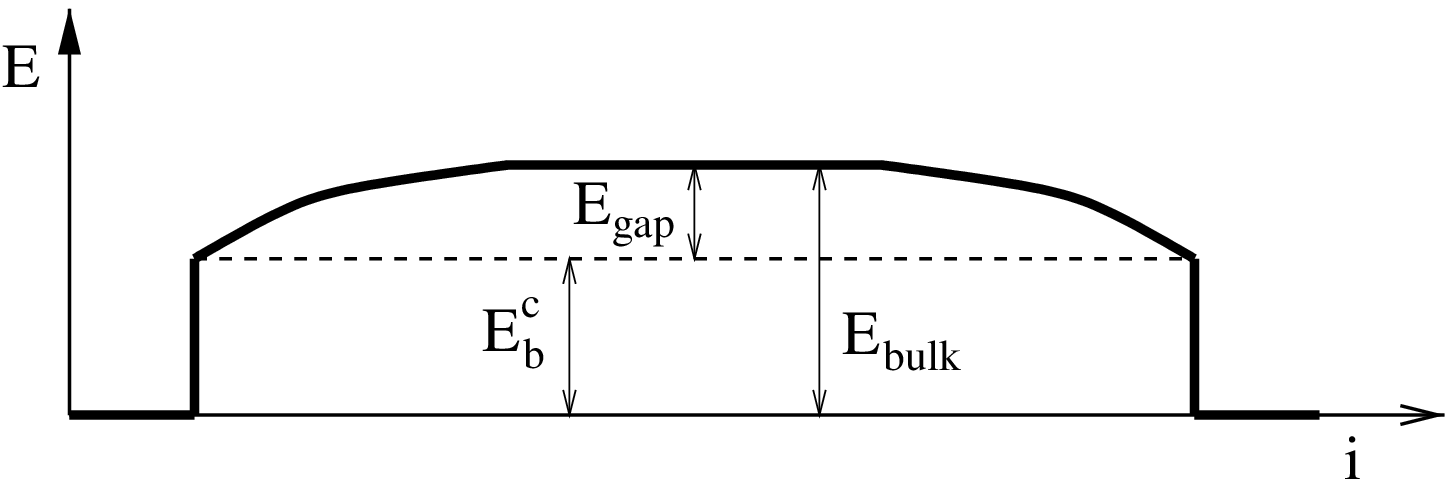}}
\vspace{1mm}
\caption{
\label{FigProfile}
Schematic drawing of the energy profile 
of a compact cluster in the improved mean field
approximation.
}
\end{figure}
}

The connection of our model to the 
experimental conditions is based on the
assumption that the tilt angle of the 
experiment tunes the ratio between
the barrier height and the difference 
of potential energies between two rows.
If the system has been prepared at 
some $\varphi_0$, we raise the 
tilt angle to $\varphi$; perturbing 
the system in this region of mixed
stability will generate an avalanche.

That is, for $\varphi > \varphi_0$ 
we have $E_b < E_b^c$. 
As the tilt angle is reduced, 
the  size of $E_b$ (measured in units of the 
potential difference)  increases, 
until it  reaches its critical value 
precisely at $\varphi_0$. Thus increasing 
$E_b$ in the model corresponds to 
lowering the tilt angle towards 
the value at which the system has been prepared
and, as such, is precisely the boundary of dynamic stability.

\FigureInText{
\begin{figure}
\epsfxsize=65mm
\centerline{\epsffile{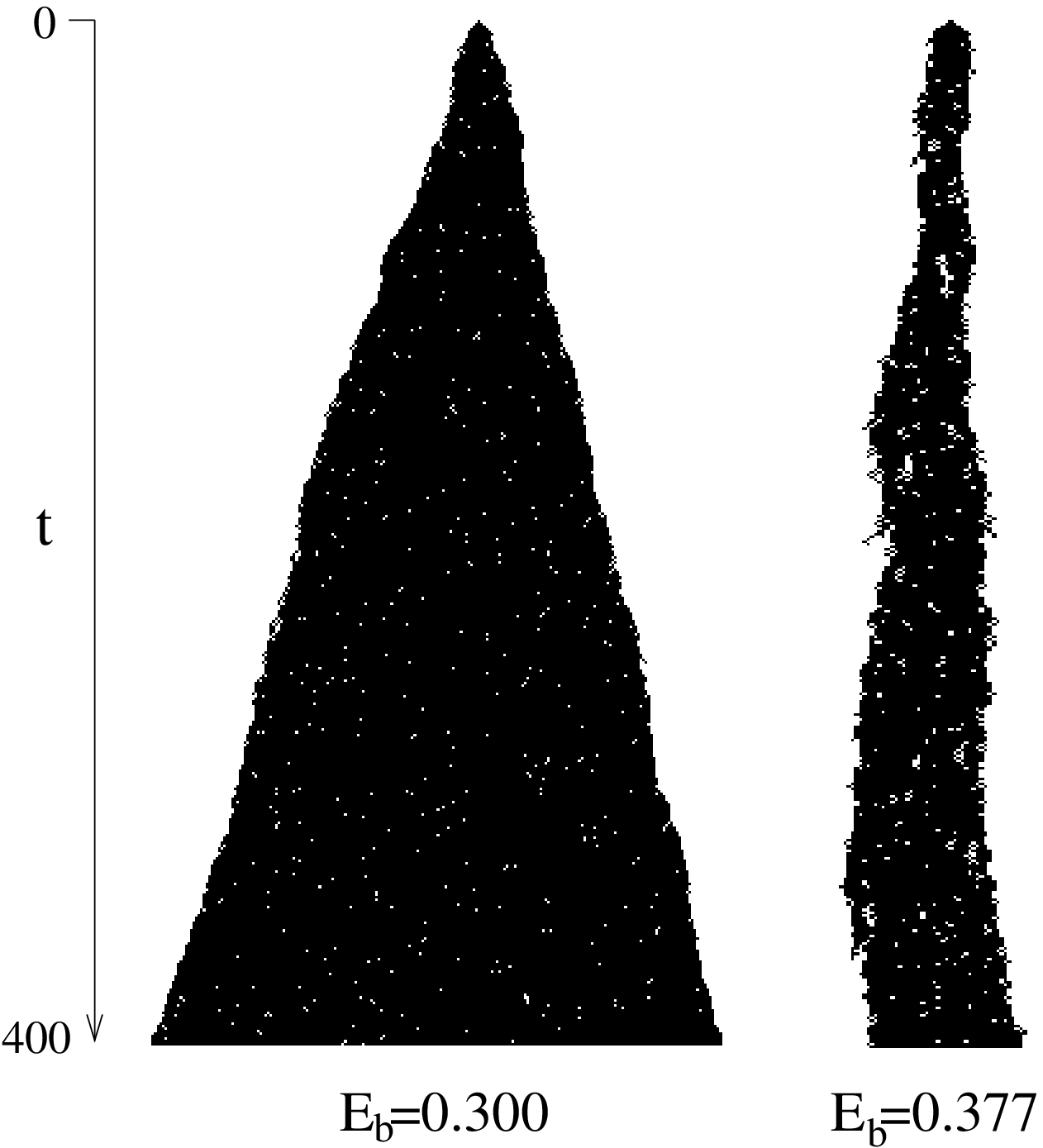}}
\vspace{2mm}
\caption{
\label{fig:avalanches}
Typical avalanches starting from a single seed
with dissipation $f=0.5$ far away and close
to criticality.
}
\end{figure}
}

Hence to reproduce the experiment we were looking for 
\begin{enumerate}
\item
fairly compact triangular regions of activation for $E_b <E_b^c$,
\item 
a varying opening angle of these triangles which should 
go to zero as $E_b$ approaches $E_b^c$ from below.
\end{enumerate}
The number of ``time steps'' that correspond 
to the DD experiment can be estimated
as the number of rows of beads from top to 
bottom of the plate, i.e. about 3000.

We simulated the model defined 
in Eqs.~(\ref{eq:DEt})-(\ref{eq:Et}) 
to check whether it is possible to reproduce 
the qualitative features of the experiment. 
Indeed we found this to be the case, as can be 
seen in Fig.~\ref{fig:avalanches}.
The two avalanches were produced for dissipation 
$f=0.5$, activating a single site at 
$t=0$, to which an initial energy of $E_0=500$ was 
assigned\footnote{Note that after less 
than 20 time steps all the initial energy has been dissipated.}.
The avalanches were compact, triangular, and with fairly straight 
edges. The edges became rough only when $E_b$
was very close to its critical value, as 
can be seen on the right
hand side of Fig.~\ref{fig:avalanches}. 
The opening angle of the active regions $\theta$ decreased as 
$E_ b$ increased towards $E_b^c$, which is shown in 
Fig.~\ref{fig:angle}. From these simulations we obtain
the estimate (see Eq.~(\ref{eq:Dnu}))
\begin{equation}
x=\nu_\parallel - \nu_\perp = 0.98(5) \simeq 1 \,.
\end{equation}
\FigureInText{
\begin{figure}
\epsfxsize=70mm
\centerline{\epsffile{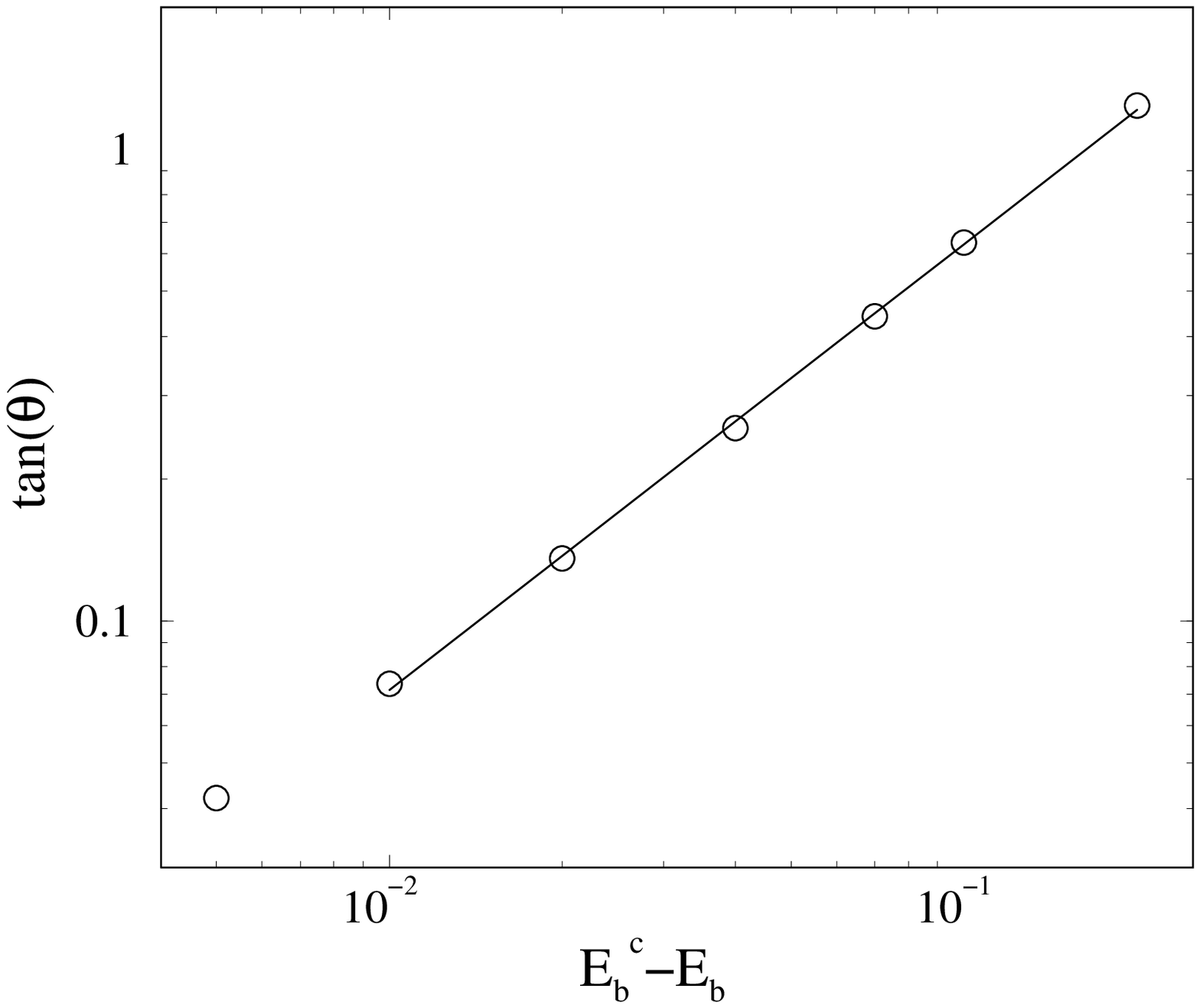}}
\caption{
\label{fig:angle}
Opening angle $\theta$ of activated triangular regions
as a function of the distance from criticality in
a double-logarithmic representation. 
}
\end{figure}
}
We predict that measuring the dependence of the avalanche 
opening angle on $\Delta \varphi$ in the experiment
should also give a linear law. 

Furthermore, the density of active sites 
in the interior of the triangular regions 
is found to be almost constant, indicating a first-order
transition. These results suggest that the transition belongs to the
CDP universality class, which is characterized by the critical 
exponents~\cite{DickmanTretyakov95}
\begin{equation}
\nu_\parallel=2\,, \qquad \nu_\perp=1 \,, \qquad \beta=0\,.
\end{equation}
These observations pose, however, a 
puzzle: since we believe that DP
is the generic situation, we would 
expect to find non-compact active regions
and DP exponents. In the following 
Section we present a careful numerical analysis  
of the critical behavior of our model 
which resolves this problem: the
exponents seen in our simulations 
(and in the experiment) should cross over to the
DP values, but only if one gets very 
deep into the critical region.  

\section{Crossover to directed percolation}

The linear law observed in Fig.~\ref{fig:angle} can be explained
by assuming compact clusters whose temporal evolution is determined 
by the fluctuations of their boundaries. The boundaries perform an 
effective random walk with a spatial bias proportional to $E_b-E_b^c$.
Therefore, the critical model should behave in the same
way as a Glauber-Ising model at zero temperature,
i.e., the transition should belong to the CDP universality class.
However, according to the DP conjecture~\cite{DPConjecture}
any continuous spreading transition from a {\em fluctuating} active phase
into a single frozen state should belong to the universality class of
directed percolation (DP), provided that the model is defined by short
range interactions without exceptional properties such as higher
symmetries or quenched randomness (see Sec. VI).
Clearly, the present model fulfills these requirements. 
It has indeed a fluctuating active state and exhibits a
phase transition into a single absorbing state which is
characterized by a positive one-component order parameter.
According to these arguments, the phase transition should
belong to the DP universality class.
%
%
%
\FigureInText{
\begin{figure}
\epsfxsize=70mm
\centerline{\epsffile{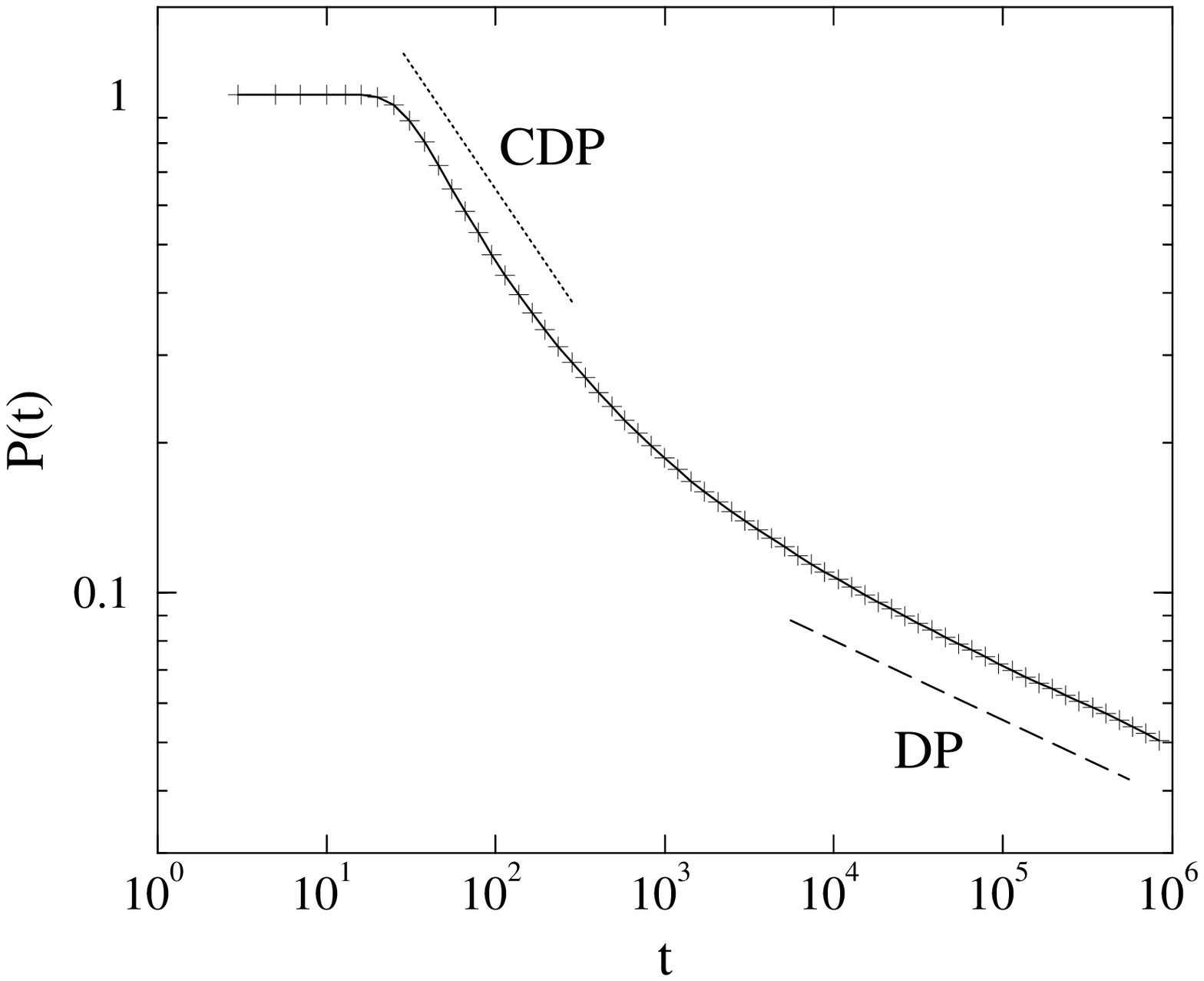}}
\caption{
Mean survival probability $P(t)$ of the toppling process
at criticality averaged over $50\,000$ independent runs.
The predicted slopes for CDP and DP are indicated by
dotted and dashed lines, respectively.
\label{FigSurv}
}
\end{figure}
}
%
%

%
%
%
%
\FigureInText{
\begin{figure}
\epsfxsize=75mm
\centerline{\epsffile{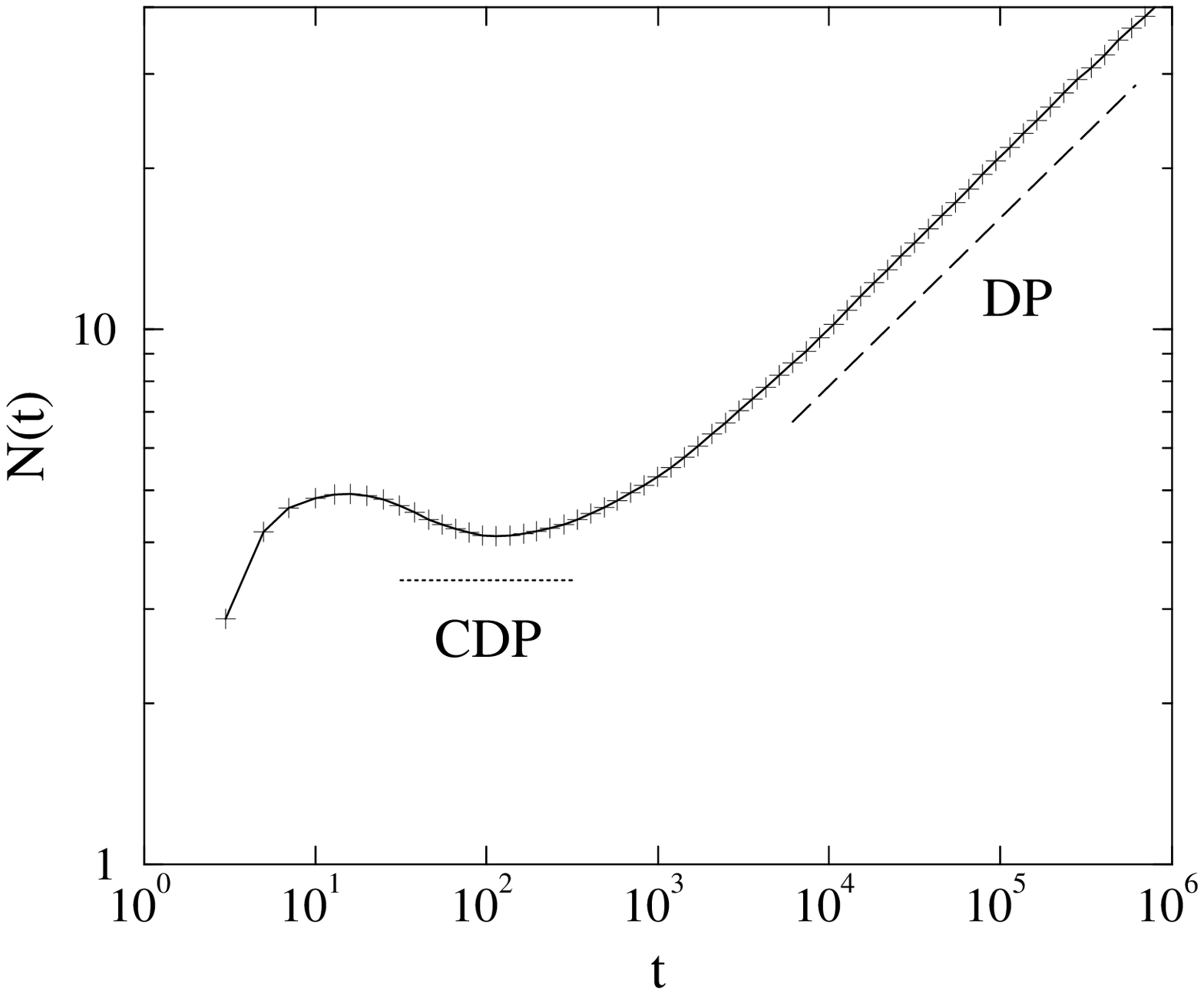}}
\caption{
Average number of active sites $N(t)$. 
The expected slopes are indicated in the same way as in 
Fig.~\ref{FigSurv}.
}
\end{figure}
}

In order to understand this apparent paradox we
perform high-precision Monte-Carlo simulations for
dissipation $f=0.5$. We employ time-dependent 
simulations~\cite{GrassbergerTorre79},
i.e., we topple a single grain in the center and
analyze the properties of the resulting cluster.
As usual for this type of simulations, we measure
the survival probability $P(t)$, the number of active
sites $N(t)$, and the mean square spreading from
the origin $R^2(t)$ averaged over the surviving
runs.  At criticality, these quantities are expected
to show an asymptotic power law behavior
\begin{equation}
P(t)\sim t^{-\delta}\,,
\qquad
N(t) \sim t^\eta\,,
\qquad
R^2(t) \sim t^{2/z}\,,
\end{equation}
where $\delta$, $\eta$, and $z$ are critical exponents
which label the universality class. In the case of CDP these
exponents are given by~\cite{DomanyKinzel,DickmanTretyakov95}
\begin{equation}
\delta=1/2\,,
\qquad
\eta=0\,,
\qquad
z=2\,,
\end{equation}
whereas DP is characterized by the exponents~\cite{Jensen}
\begin{equation}
\delta=0.1595\,,
\qquad
\eta=0.3137\,,
\qquad
z=1.5807\,.
\end{equation}
%
%
%
\FigureInText{
\begin{figure}
\epsfxsize=75mm
\centerline{\epsffile{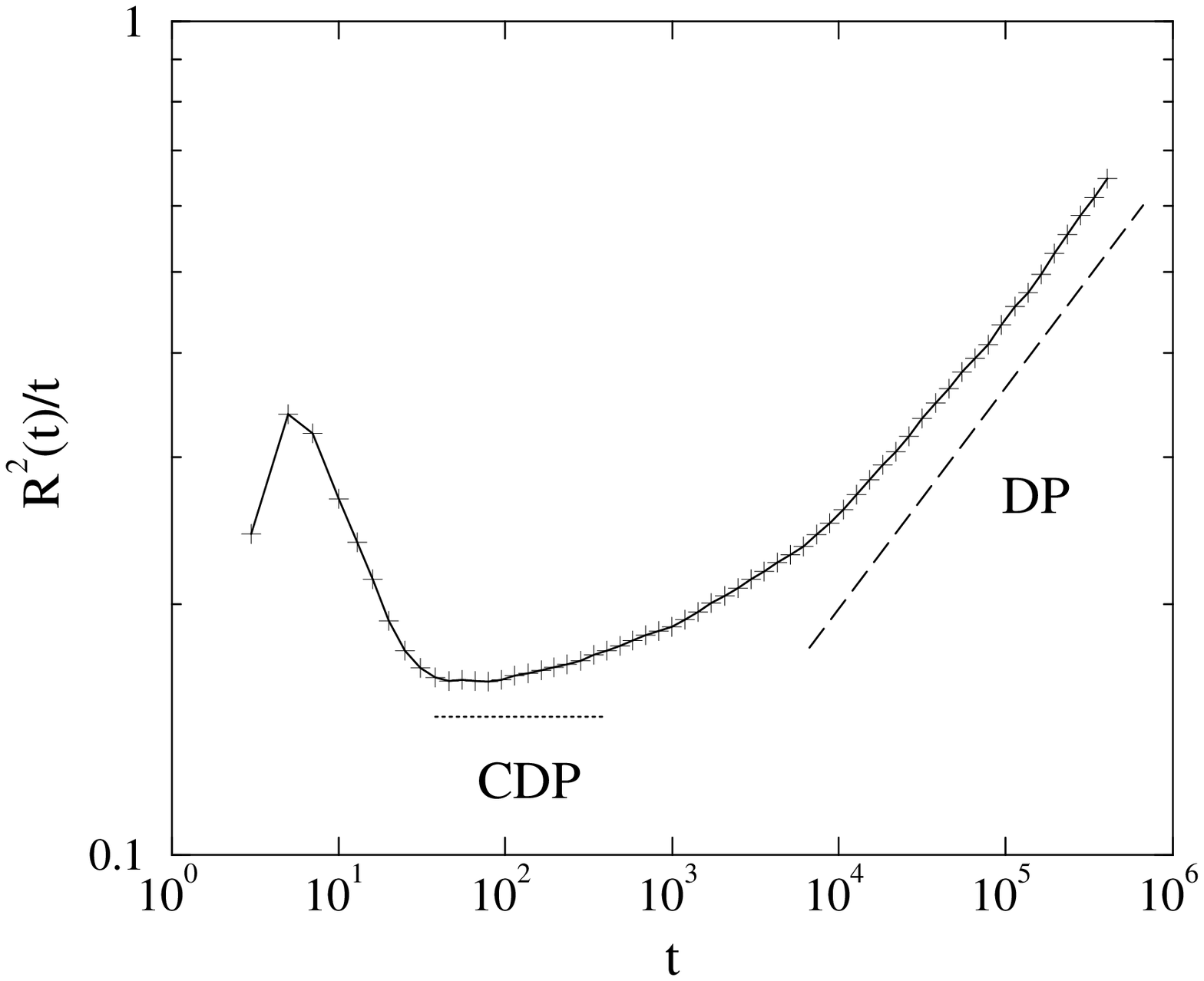}}
\caption{
Mean square spreading from the origin averaged over surviving
runs. In order to demonstrate the crossover from CDP to DP
we divided $R^2(t)$ by $t$. The expected slopes $(2-z)/z$
are indicated by dotted and dashed lines.
\label{FigSpread}
}
\end{figure}
}
In order to eliminate finite-size effects, we use a dynamically
generated lattice adjusted to the actual size of the cluster.
Moreover, we observe that the initial non-universal transient
is minimal if an excitation energy $E_0 \simeq 15$ is used.
Detecting deviations from power-law behavior in the long-time limit
we estimate the critical energy by $E_b^c=0.385997(5)$.
Our numerical results (obtained from simulations at the critical point) 
are shown in Figs.~\ref{FigSurv}-\ref{FigSpread}.
In all measurements we observe different temporal regimes:
\begin{enumerate}
\item
During the first few time steps,
the activation energy is distributed to the nearest
neighbors whereby the cluster grows at maximal speed.
Therefore, the survival probability $P(t)$ is $1$  and
the particle number $N(t)$ grows linearly.

\item
In the intermediate regime, which extends up to a few
hundred time steps, the inactive islands within the
cluster are not yet able to break up the cluster into 
separate parts. Thus, the cluster can be considered
as being compact and the temporal evolution is
governed by a random walk of its boundaries. In this
regime we observe a power-law behavior with CDP exponents
(indicated by dotted lines in
Figs.~\ref{FigSurv}-\ref{FigSpread}).

\item The intermediate regime is followed by a long
crossover from CDP to DP extending over almost two decades up to
more than $10^4$ time 
steps\footnote{Note that the crossover
in the present model is different 
from the one studied in~\cite{MDH96},
where inhomogeneous interactions at the cluster's 
boundaries were assumed.}.

\item Finally the system enters an asymptotic DP regime
(indicated by dashed lines in Figs.~\ref{FigSurv}-\ref{FigSpread}).
\end{enumerate}

%
%
%
\FigureInText{
\begin{figure}
\epsfxsize=80mm
\centerline{\epsffile{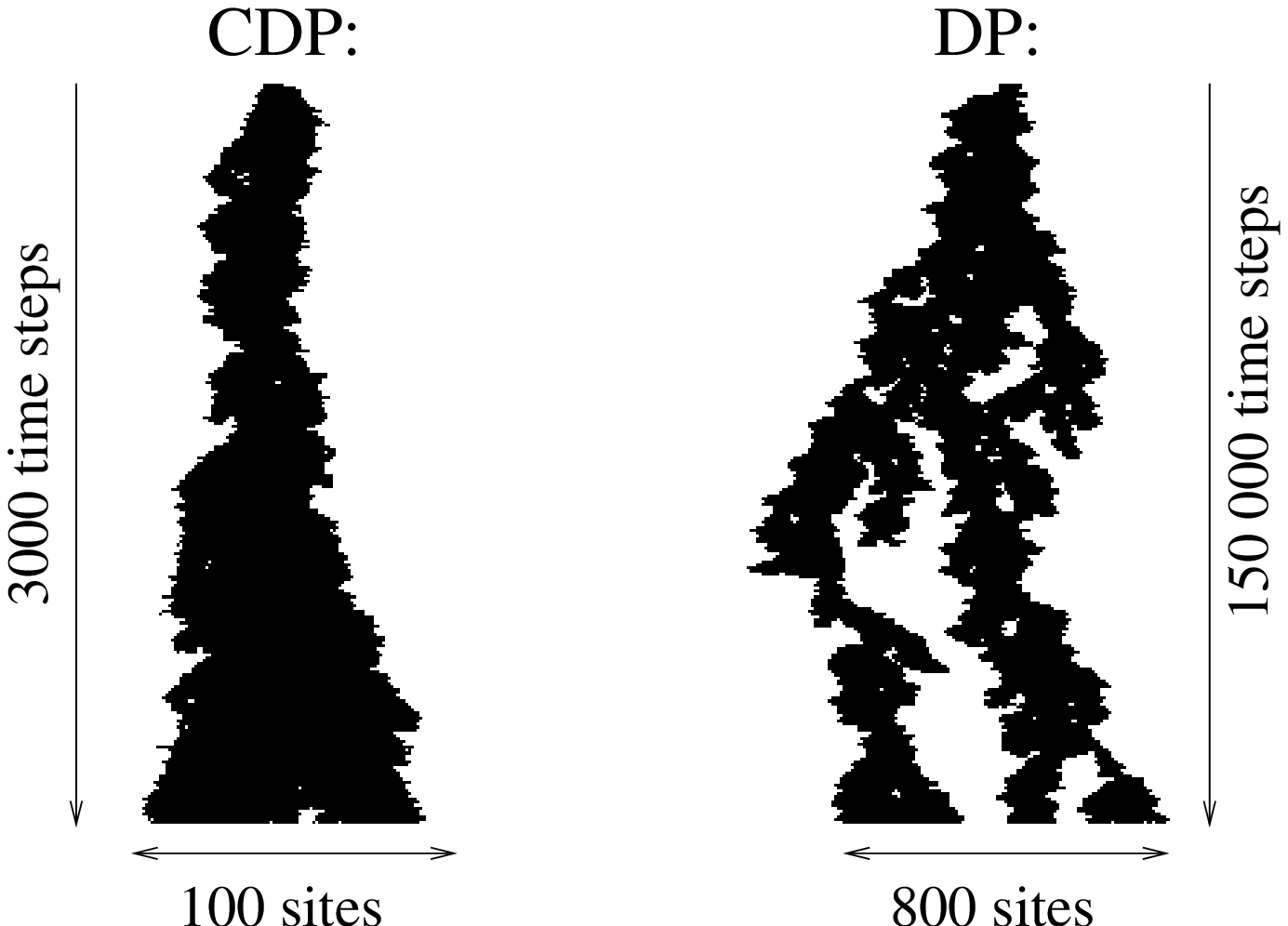}}
\vspace{2mm}
\caption{
Typical clusters generated at 
criticality on small and large scales,
illustrating the crossover from CDP to DP.
\label{FigDemoCrossover}
}
\end{figure}
}

\noindent
The crossover from CDP to DP is illustrated in
Fig.~\ref{FigDemoCrossover}.
Two avalanches are plotted on different scales. The 
left one represents a typical avalanche within the first
few thousand time steps. As can be seen, the
cluster appears to be compact on a lateral scale up to 100 lattice sites.
However, as shown in the right panel of Fig.~\ref{FigDemoCrossover},
after a very long time the cluster breaks up into several branches. The
right hand figure shows a typical cluster on a scale of $150\,000$ time
steps, where the branches still have a 
certain characteristic thickness. Going to even
larger scales the width of the branches becomes 
irrelevant and we obtain the typical
patterns of critical DP clusters.

In comparison with ordinary DP lattice models, in the
present model the observed crossover is unusually slow. 
This due to short-range correlations 
between active sites leading to active branches with 
a certain typical thickness $\xi_{act}$.
In ordinary DP lattice models the average size of active
branches is of the order of a few lattice spacings. 
In the present case, however, we find a much larger 
value $\xi_{act} \approx 20$.

%
%
%
\FigureInText{
\begin{figure}
\epsfxsize=85mm
\centerline{\epsffile{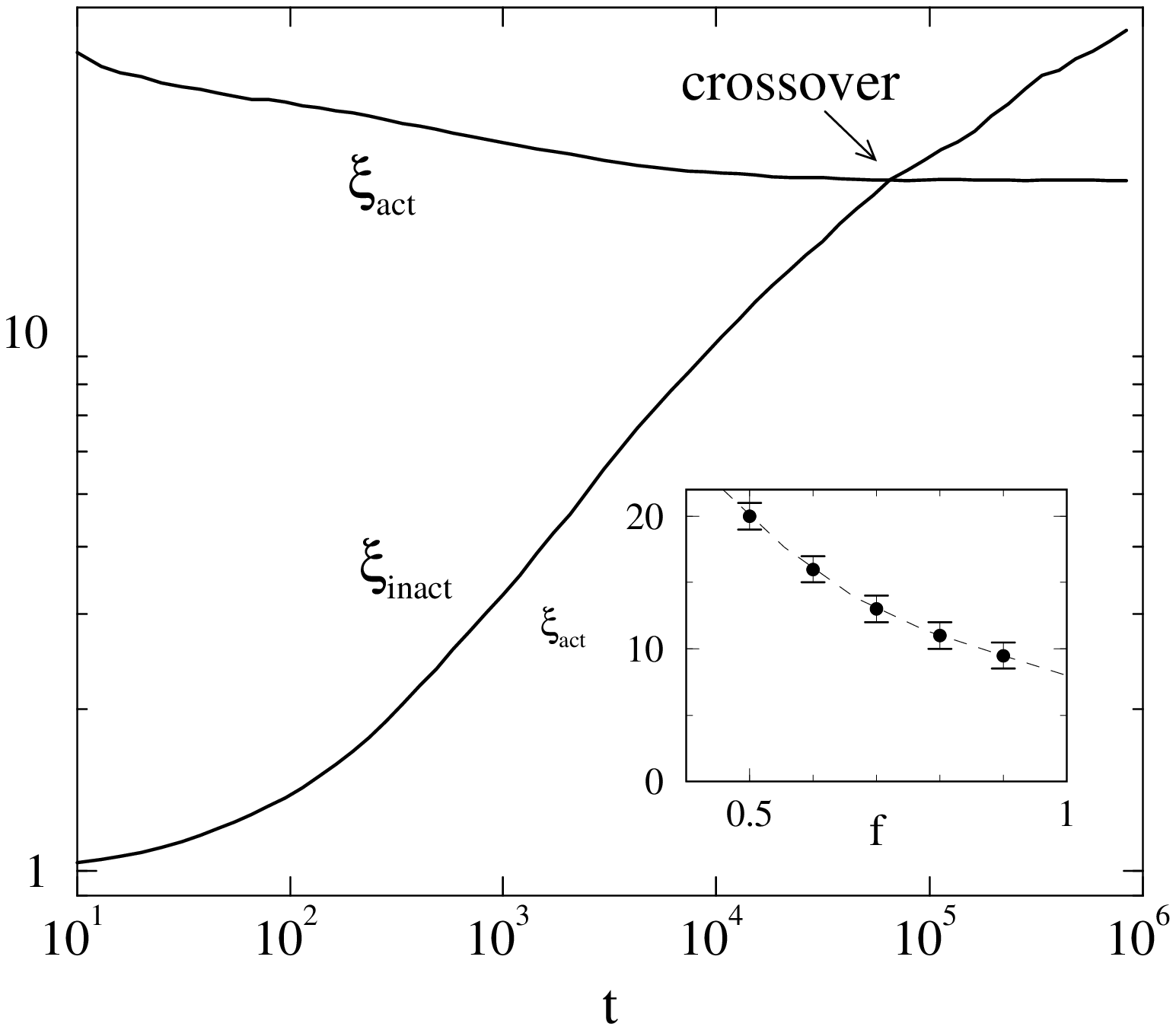}}
\caption{
Mean sizes of active and inactive regions as a
function of time, starting from homogeneous initial
conditions with dissipation $f=0.5$ (see text).
The inset shows the saturation value of $\xi_{act}$
as a function of the dissipation $f$.
\label{FigIslands}
}
\end{figure}
}

Based on this observation, the typical 
crossover time~$t_c$ can be approximated
as follows. In order to cross over to DP, the average 
size of  {\it inactive} regions between neighboring branches
$\xi_{inact}$ has to become larger than the thickness
of the branches $\xi_{act}$. 
In Fig.~\ref{FigIslands} we plot both quantities as
a function of time at criticality, using
a lattice with $N=2^{14}$ sites and homogeneous
initial conditions $E_i^{t=0}=2$.
Initially $\xi_{act}=N$ and $\xi_{inact}=0$.
As time evolves, the average size of active branches decreases
and saturates at a constant value $\xi_{act} \approx 20$.
However, the average size 
of inactive regions $\xi_{inact}$
continues to grow and exceeds $\xi_{act}$ at 
time $t_c \approx 10^5$. As can be seen, this 
provides a good estimate of the typical 
time where the critical behavior 
of the system crosses over to DP.

In order to observe the crossover experimentally, it
would be interesting to know how the crossover time $t_c$
can be reduced. To this end we measure $\xi_{act}$ 
for several values of the dissipation $f$ 
(see inset of Fig.~\ref{FigDemoCrossover}).
It turns out that by increasing $f$ the typical size
of active branches can be decreased 
down to $10$ lattice spacings. Consequently, 
the crossover time can be reduced by more 
than one decade. Hence, for an experimental 
verification of DP, systems with high 
dissipation are more appropriate.

The influence of the dissipation can easily be explained
within the improved mean field approximation of Sect.~IV.
Clearly, the stability of a cluster against breakup into
several branches by fluctuations depends on the
energy gap $E_{gap}=E_{bulk}-E_c$. As can easily be verified, 
this energy difference (and therewith the stability of 
compact clusters) decreases with increasing 
dissipation $f$, explaining the observed $f$-dependence.

\section{The effect of randomness}

The above model describes the physics of flowing sand
in a highly idealized manner. In particular, it ignores
the fact that spreading avalanches may be subjected to 
frozen disorder. For example, irregularities of the plate 
and the velvet cloth could lead to quenched randomness in the 
equations of motion. Moreover, the system prepares itself
in an initial state which is not fully homogeneous.
Thus, we have to address the question to what extent 
quenched randomness will affect the expected crossover to DP.

Certain types of quenched disorder are known to 
change the critical behavior of DP. For example, 
Moreira and Dickman studied the diluted contact
process with {\em spatially} quenched disorder~\cite{MoreiraDickman96}. 
Even for small amplitudes quenched randomness was found to destroy 
the  DP transition, turning algebraic into logarithmic laws. 
Janssen~\cite{Janssen97} confirmed and substantiated these findings 
by a field-theoretic analysis. Recently Cafiero et al.~\cite{CGM98}
mapped DP with spatially quenched disorder onto a non-markovian
process with memory exhibiting the same nonuniversal properties.
The memory is due to the formation of bound states of particles in
those regions where the percolation probability is very high.
As shown by Webman et al., these bound states give rise to a 
glassy phase separating active and inactive parts of the phase
diagram~\cite{WACH98}. Similar nonuniversal properties were 
also observed  in DP processes with {\em temporally} 
quenched disorder~\cite{Jensen96}.

In all cases investigated so far, quenched disorder 
destroys the DP transition. However, the disorder in the 
DD experiment is different in nature. Clearly, it is neither 
spatially nor temporally quenched, rather it depends on both 
space and time. On the level of our model we may think of
randomly varying energy barriers
\begin{equation}
E_b \rightarrow E_b + A \eta(x,t)\,,
\end{equation}
where the amplitude $A$ controls the intensity of disorder.
Here $\eta(x,t)$ is a white Gaussian
noise specified by the correlations
\begin{equation}
\overline{\eta(x,t)\eta(x',t')} =
\delta^d(x-x')\delta(t-t') \,,
\end{equation}
where $d=1$ denotes the spatial dimension.
In the standard situation of
quenched noise of this type $\eta(x,t)$ is kept fixed 
while the experiment is repeated and the quantities
under investigation are averaged over many independent
avalanches. Yet in the DD experiment, the situation is different.
Here once the sand has been poured, a particular realization of the random
variables has been selected. However, there is no process to repeat the
experiment over and over again with a fixed $\eta(x,t)$. Rather, after each
avalanche the system is prepared again (by pouring sand or by starting an
avalanche elsewhere). Hence the averaging process is done simultaneously 
over the $\eta(x,t)$ {\it and} the stochastic dynamic process 
that generates the avalanches. This type of averaging is of 
the annealed type and therefore less likely to alter the critical 
behavior than its quenched version. 

In order to find out whether fully quenched disorder
affects the asymptotic critical behavior of DP, we simulated a
directed bond percolation process with randomly distributed
bond probabilities between $p^*$ and $1$.  
For $p^*=0.289(1)$, we find asymptotic power laws with 
DP exponents, indicating that the transition 
is not affected by spatio-temporally quenched noise.Therefore,
we expect the same to be true in the case of annealed 
disorder in our model for flowing sand.

To support this point of view, we study the case of quenched randomness
in the 
DP Langevin equation~\cite{Janssen81}
\begin{eqnarray}
\label{LangevinEquation}
\partial_t\rho(x,t) &=& a\rho(x,t) - g\rho^2(x,t) + 
\nonumber\\
&&D\nabla \rho(x,t) + 
\Gamma \sqrt{\rho(x,t)} \xi(x,t) \,,
\end{eqnarray}
where $\rho(x,t)$ is the particle density 
and $a$ represents the percolation probability. 
$\xi(x,t)$ is a Gaussian white noise which represents 
the intrinsic randomness of the DP process. 
At the critical dimension $d=4$, where
fluctuations start to contribute, the
Langevin equation~(\ref{LangevinEquation}) is 
invariant under scaling transformations
$x\rightarrow bx$, $t \rightarrow b^2 t$, and 
$\rho \rightarrow b^{-2}\rho$. 

In order to include spatio-temporally 
quenched randomness, we allow for small variations
of $a$, i.e., we add the term 
$$
A \, \rho(x,t) \, \eta(x,t)
$$
on the right hand side of Eq.~(\ref{LangevinEquation}). 
However, as can be shown by simple dimensional analysis,
this term is {\em irrelevant} in $d=4$ dimensions, i.e.,
it decreases and eventually vanishes
under scaling transformations. This observation
strongly supports the result that the DP transition in our model
is indeed not affected by quenched randomness.

We emphasize that the irrelevance of quenched randomness
in our model is due to the special role of 'time' which 
coincides with the vertical coordinate of the plane. That is,
for each time step the stochastic processes take place
in a different random environment. To that extent 
the DD experiment differs from other DP-related 
experiments such as catalytic reactions where spatially
quenched disorder affects the critical behavior.

\section{Conclusions}

We introduced a simple model for 
flowing sand on an inclined plane. The model is inspired by 
recent experiments and reproduces some of the observed
features. In contrast to the experiment, which prepares 
itself in a self-organized critical state, our model needs
to be tuned to a critical point by varying the energy
barrier $E_b$. At criticality the system undergoes a 
nonequilibrium phase transition from an inactive (dry) 
phase with finite avalanches to an active (wet)
phase where the mean size of avalanches diverges. Analyzing the
critical behavior near the transition, we obtained the
following results:

\begin{enumerate}

\item
On short scales, i.e., on scales considered in the DD experiment,
the model reproduces the experimentally observed
triangular compact avalanches. In the active phase their
opening angle $\theta$ is predicted to vary 
linearly\footnote{Note added after submission: This
prediction has to be compared with the model proposed
by Bouchaud {\it et al.} \cite{BouchaudCates98} 
which predicts the exponent $x=1/2$.} 
with $\Delta \varphi$.

\item
On very large scales the critical behavior of the
model crosses over to ordinary DP. Thus, the DD experiment 
could serve as a first physical realization
of directed percolation. 
Crossover to DP is seen in the model after about $10^4$ time steps,
whereas the DD experiment stops at about 3000 steps (i.e. rows of beads).
Hence in order to observe the crossover in the experiment, 
larger system sizes and/or smaller beads 
would be required.

\item
We have shown that quenched randomness with short-range 
correlations due to irregularities in the experiment 
should not affect the asymptotic critical behavior.

\item
The typical time needed to cross over to
DP is found to decrease with increasing dissipation.

\end{enumerate}

Thus, in order to create experimental conditions favoring
a crossover to DP, we suggest to use small glass
beads, large system sizes, and an initial angle $\varphi_0$
where the dissipation of energy per toppling grain is
maximal. For physical reasons we would expect the dissipation 
to be maximal for small angles $\varphi_0$, but this has to
be verified in the actual experiment.

As a necessary precondition for a crossover to DP, 
compact clusters must be able to split up into several
branches, as illustrated in Fig.~\ref{FigDemoCrossover}.
Thus, before measuring critical exponents, this feature 
has to be tested experimentally.
To this end the DD experiment should 
be performed repeatedly at the critical tilt 
$\varphi=\varphi_0$. In most cases the avalanches will be
small and compact. However, large avalanches, 
reaching the bottom of the plate, will sometimes be generated.
If these avalanches are non-compact (consisting of 
several branches) we expect the asymptotic critical
behavior to be described by DP. Only then it is
worthwhile to optimize the experimental setup and
to measure the critical exponents quantitatively.

\vspace{4mm}
\noindent
{\em Acknowledgements}\\
AJD wishes to thank the kind hospitality of the Weizmann Institute 
and acknowledges financial support from the CICPB and the UNLP,
Argentina and from the Weizmann Institute.
HH thanks the Weizmann Institute and the 
Einstein Center for hospitality and financial support.
ED thanks the Germany-Israel Science Foundation (GIF) for partial
support, B. Derrida for some most helpful initial insights
and A. Daerr for communicating his results at an early stage.

\vspace{4mm}



\TextEndsHere
\newpage
\centerline{FIGURES CAPTIONS}
\vspace{5mm}

\noindent Fig. 1:
Schematic stability diagram of the DD-experiment.
A layer of thickness $h$ is dynamically stable
below a certain threshold $h_d(\varphi)$ (solid
line). Due to friction forces non-moving layers remain
stable in the mixed region below static stability 
limit $h_s(\varphi)$ (dashed line).
In the present work we investigate the properties
in the vicinity of the dynamic phase transition
line, as indicated by the arrows.\\

\noindent Fig. 2:
The top layer of sand in an effective washboard potential.\\

\noindent Fig. 3:
Energy transfer between grains on a square lattice.\\

\noindent Fig. 4:
Phase diagram of the model for flowing sand. 
The full line represents the phase transition 
line. The mean-field approximations
of Eqs.~(\ref{SimpleMeanField}) 
and~(\ref{ImprovedMeanField}) are
shown as dotted and dashed lines, 
respectively. \\

\noindent Fig. 5:
Schematic drawing of the energy profile 
of a compact cluster in the improved mean field
approximation.\\

\noindent Fig. 6:
Typical avalanches starting from a single seed
with dissipation $f=0.5$ far away and close
to criticality.\\

\noindent Fig. 7:
Opening angle $\theta$ of activated triangular regions
as a function of the distance from criticality in
a double-logarithmic representation. \\

\noindent Fig. 8:
Mean survival probability $P(t)$ of the toppling process
at criticality averaged over $50\,000$ independent runs.
The predicted slopes for CDP and DP are indicated by
dotted and dashed lines, respectively.\\

\noindent Fig. 9:
Average number of active sites $N(t)$. 
The expected slopes are indicated in the same way as in 
Fig.~\ref{FigSurv}.\\

\noindent Fig. 10:
Mean square spreading from the origin averaged over surviving
runs. In order to demonstrate the crossover from CDP to DP
we divided $R^2(t)$ by $t$. The expected slopes $(2-z)/z$
are indicated by dotted and dashed lines.\\

\noindent Fig. 11:
Typical clusters generated at 
criticality on small and large scales,
illustrating the crossover from CDP to DP.\\

\noindent Fig. 12:
Mean sizes of active and inactive regions as a
function of time, starting from homogeneous initial
conditions with dissipation $f=0.5$ (see text).
The inset shows the saturation value of $\xi_{act}$
as a function of the dissipation $f$.\\

\newpage
\begin{figure}
\epsfxsize=130mm
\centerline{\epsffile{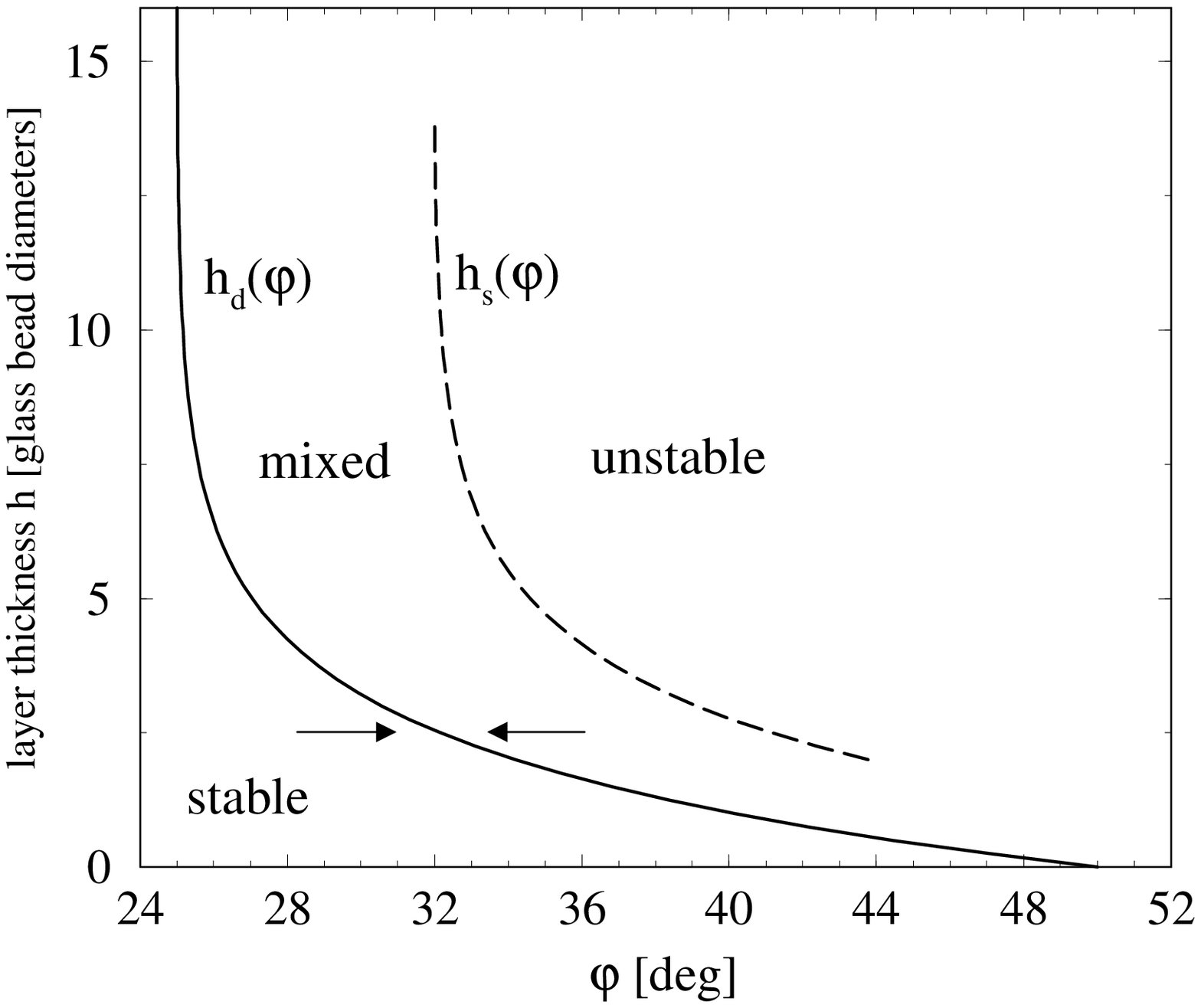}}
\caption{\label{FigExperiment}}
\end{figure}

\vspace{20mm}
\begin{figure}
\epsfxsize=100mm
\centerline{\epsffile{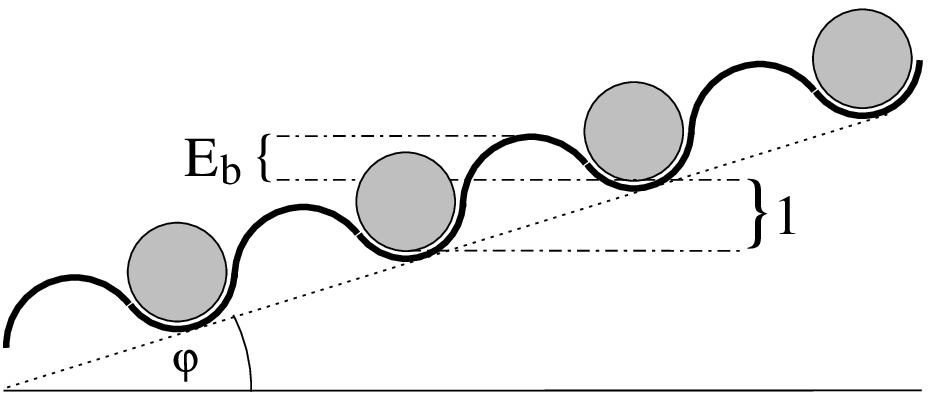}}
\caption{\label{FigPotential}}
\end{figure}

\begin{figure}
\epsfxsize=100mm
\centerline{\epsffile{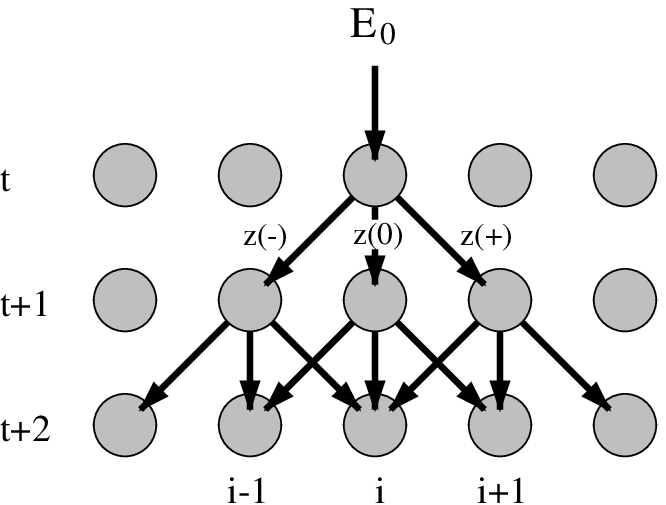}}
\caption{\label{FigSquareLattice}}
\end{figure}

\vspace{20mm}
\begin{figure}
\epsfxsize=120mm
\centerline{\epsffile{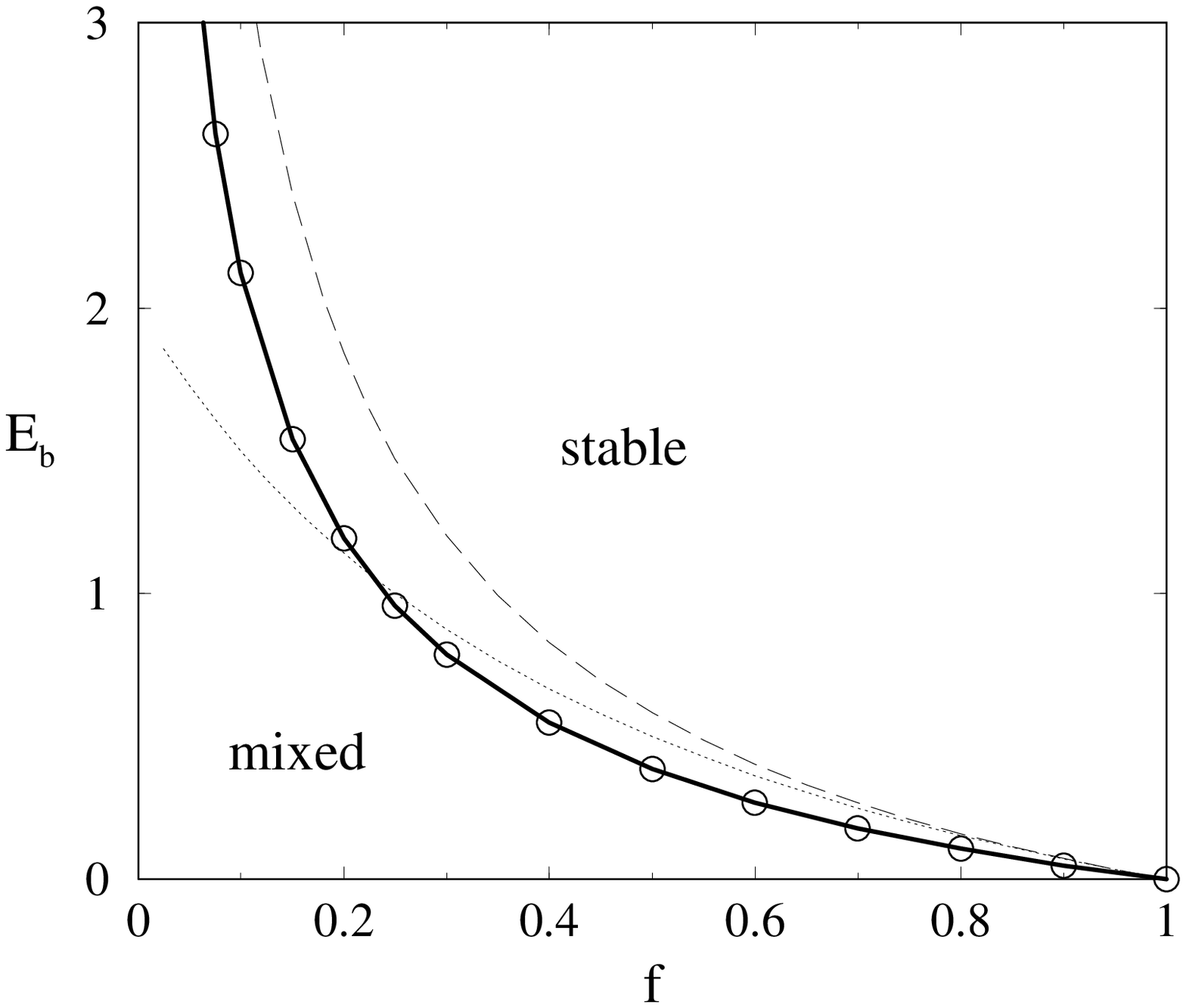}}
\caption{\label{FigPhaseTrans}}
\end{figure}

\begin{figure}
\epsfxsize=120mm
\centerline{\epsffile{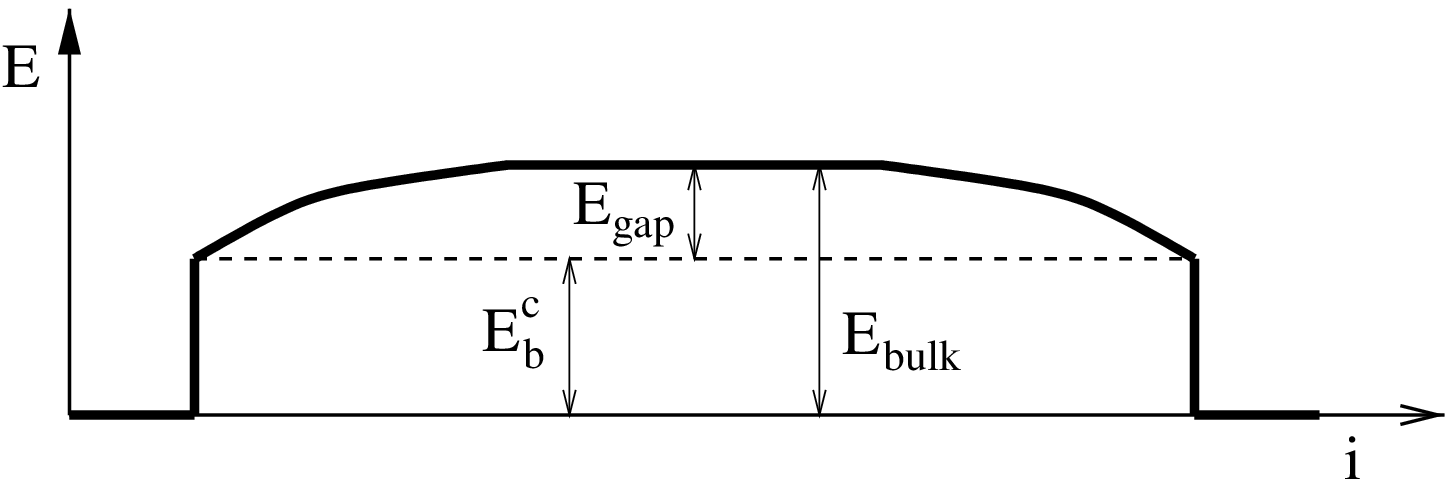}}
\caption{\label{FigProfile}}
\end{figure}

\vspace{20mm}
\begin{figure}
\epsfxsize=100mm
\centerline{\epsffile{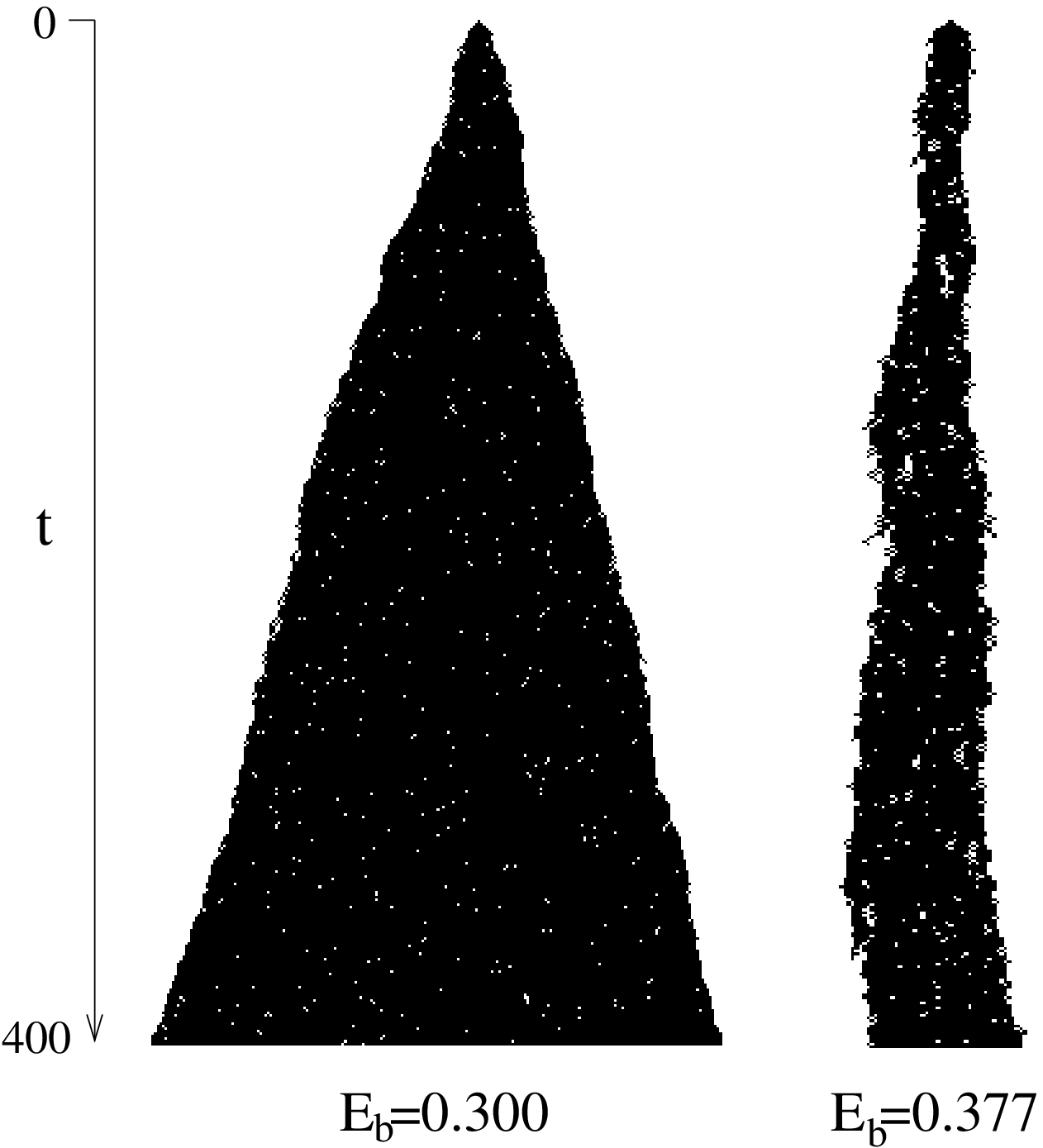}}
\caption{\label{fig:avalanches}}
\end{figure}

\begin{figure}
\epsfxsize=120mm
\centerline{\epsffile{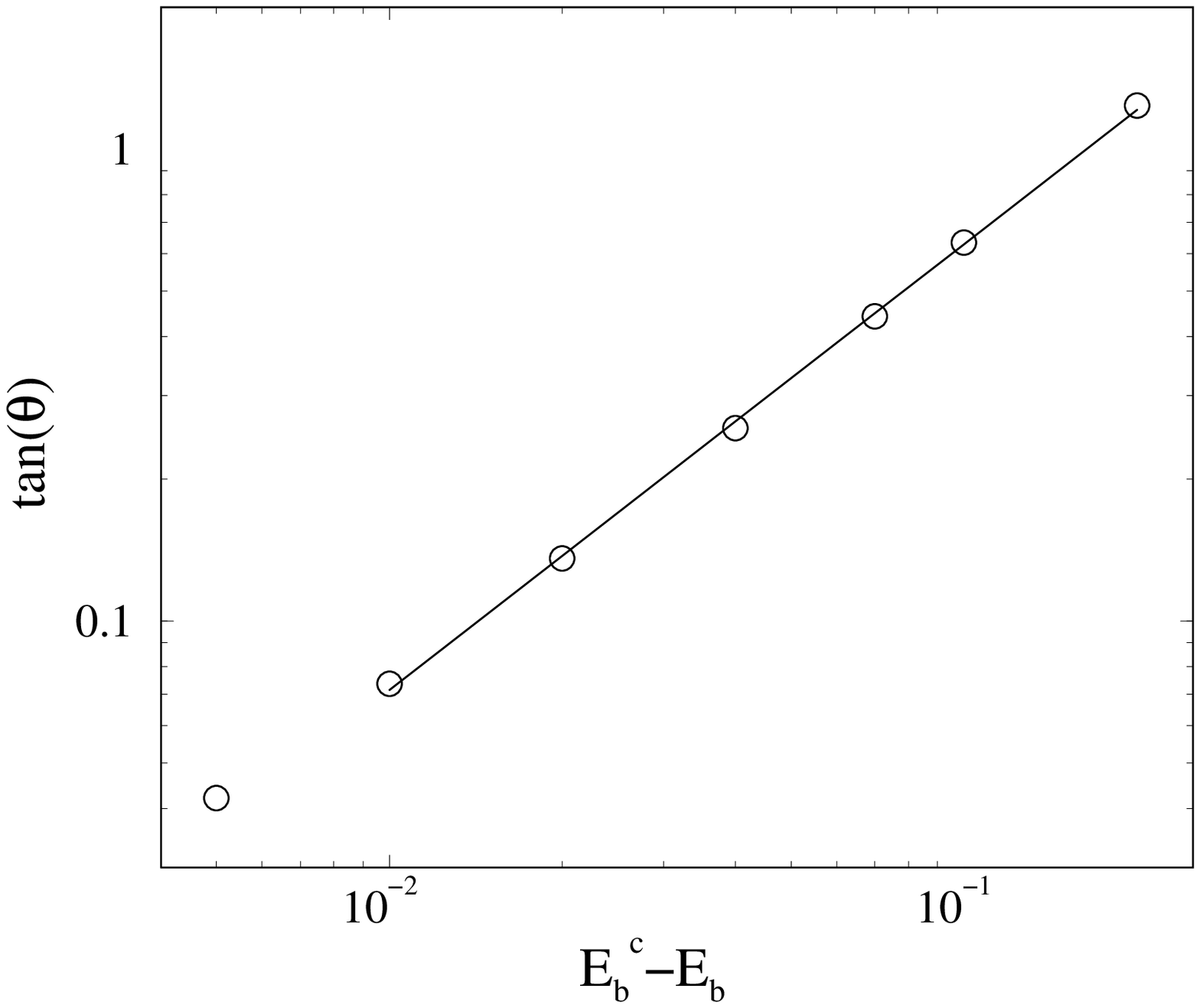}}
\caption{\label{fig:angle}}
\end{figure}

\begin{figure}
\epsfxsize=120mm
\centerline{\epsffile{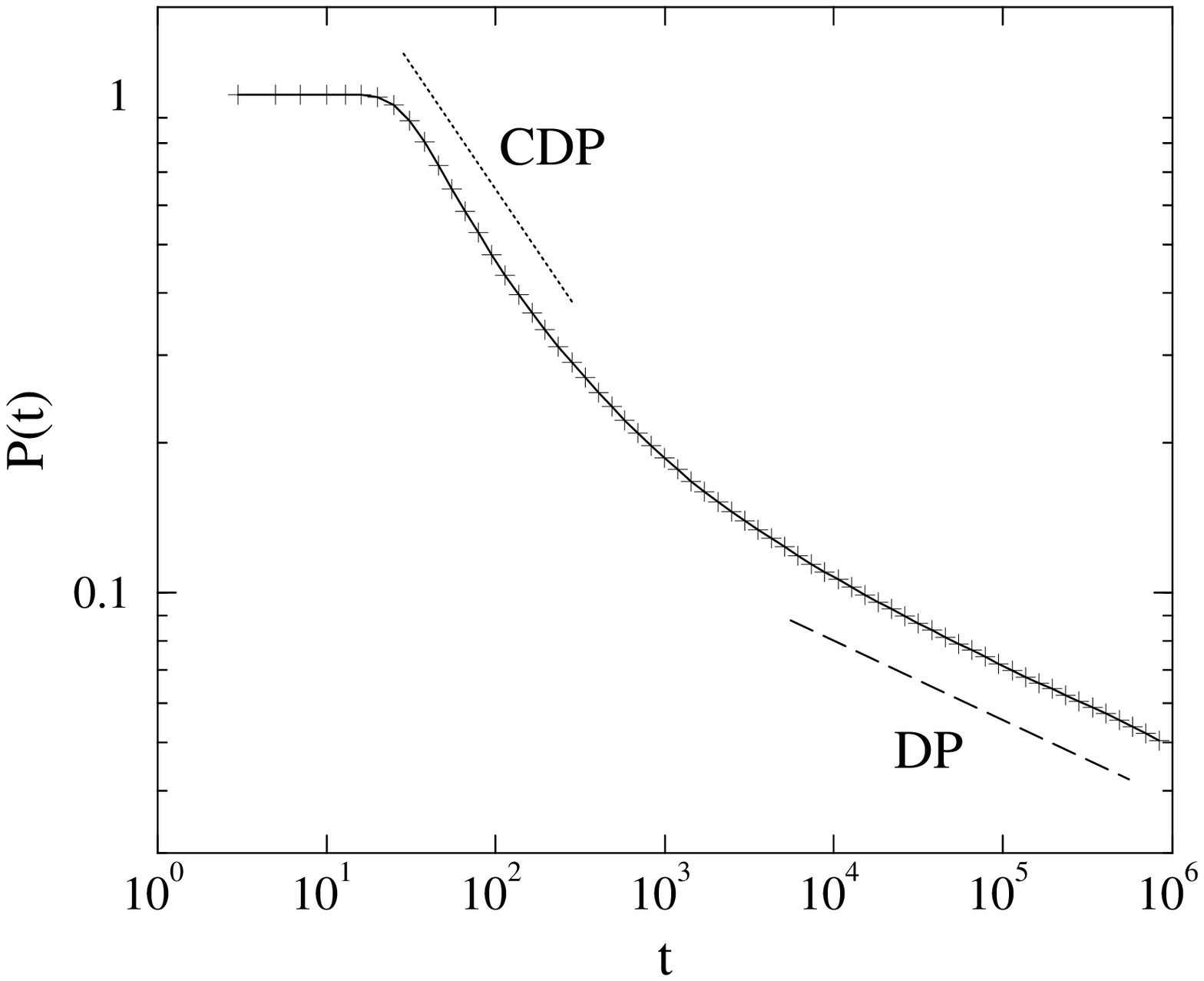}}
\caption{\label{FigSurv}}
\end{figure}

\begin{figure}
\epsfxsize=120mm
\centerline{\epsffile{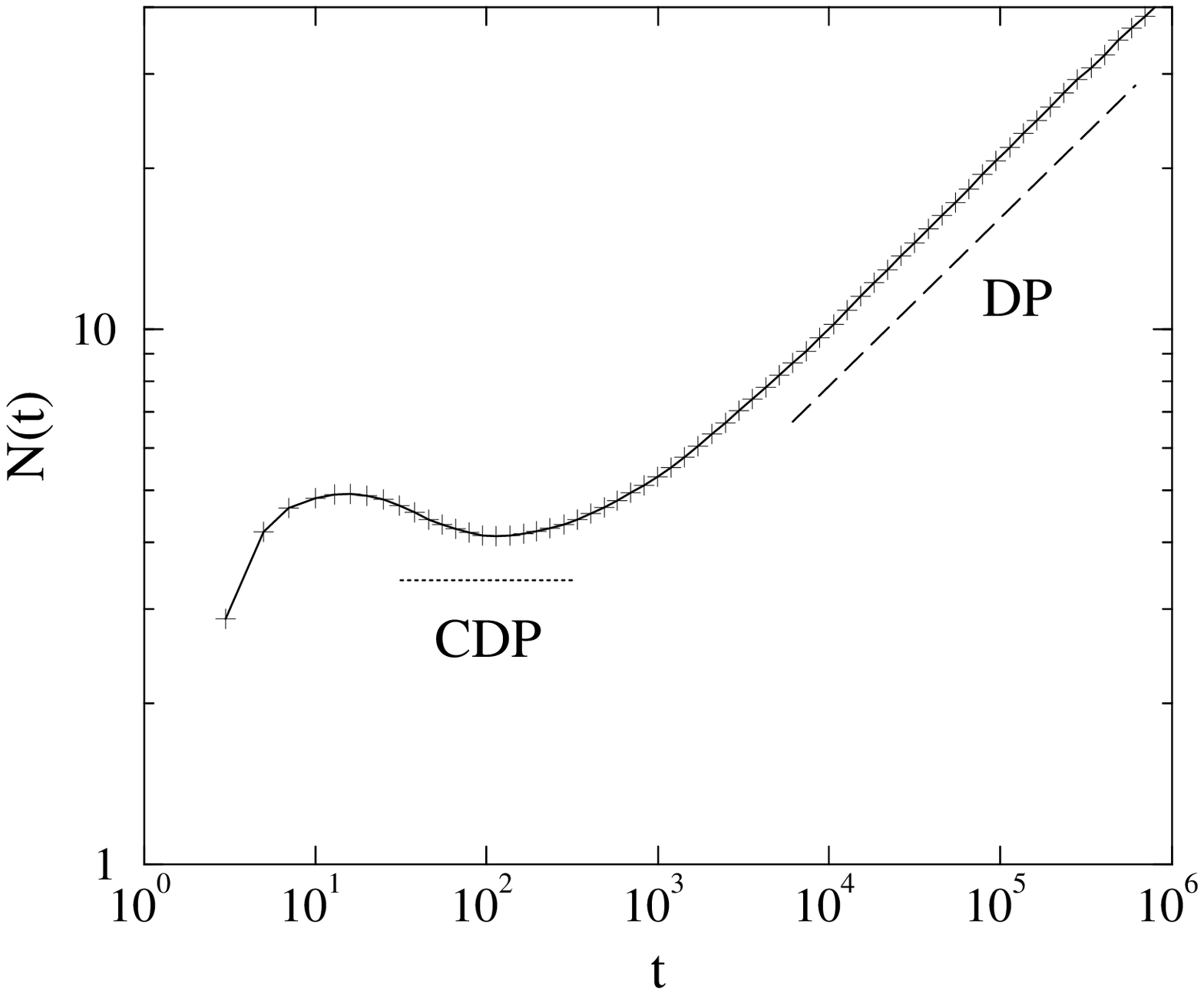}}
\caption{\label{FigMass}}
\end{figure}

\begin{figure}
\epsfxsize=120mm
\centerline{\epsffile{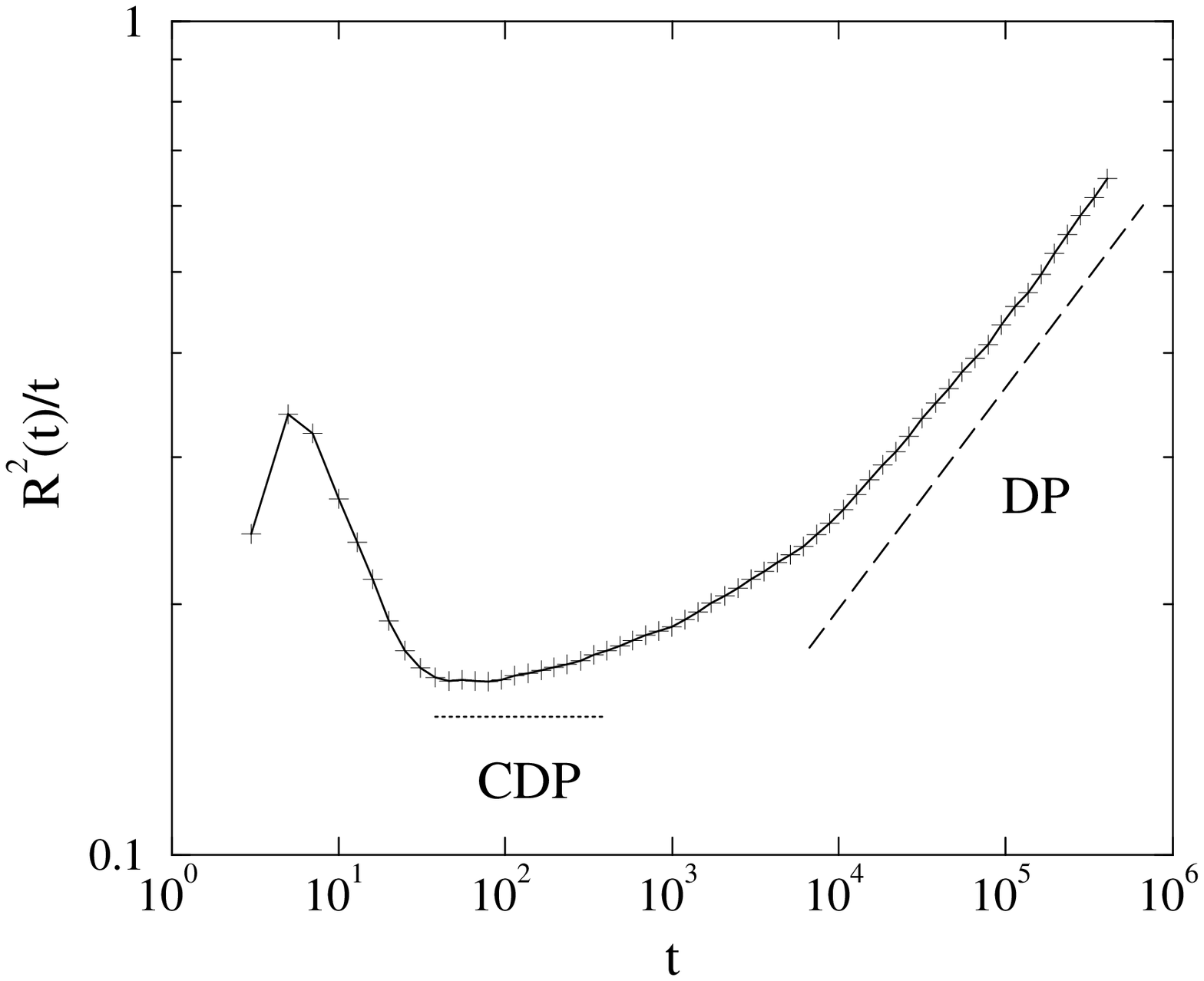}}
\caption{\label{FigSpread}}
\end{figure}

\begin{figure}
\epsfxsize=130mm
\centerline{\epsffile{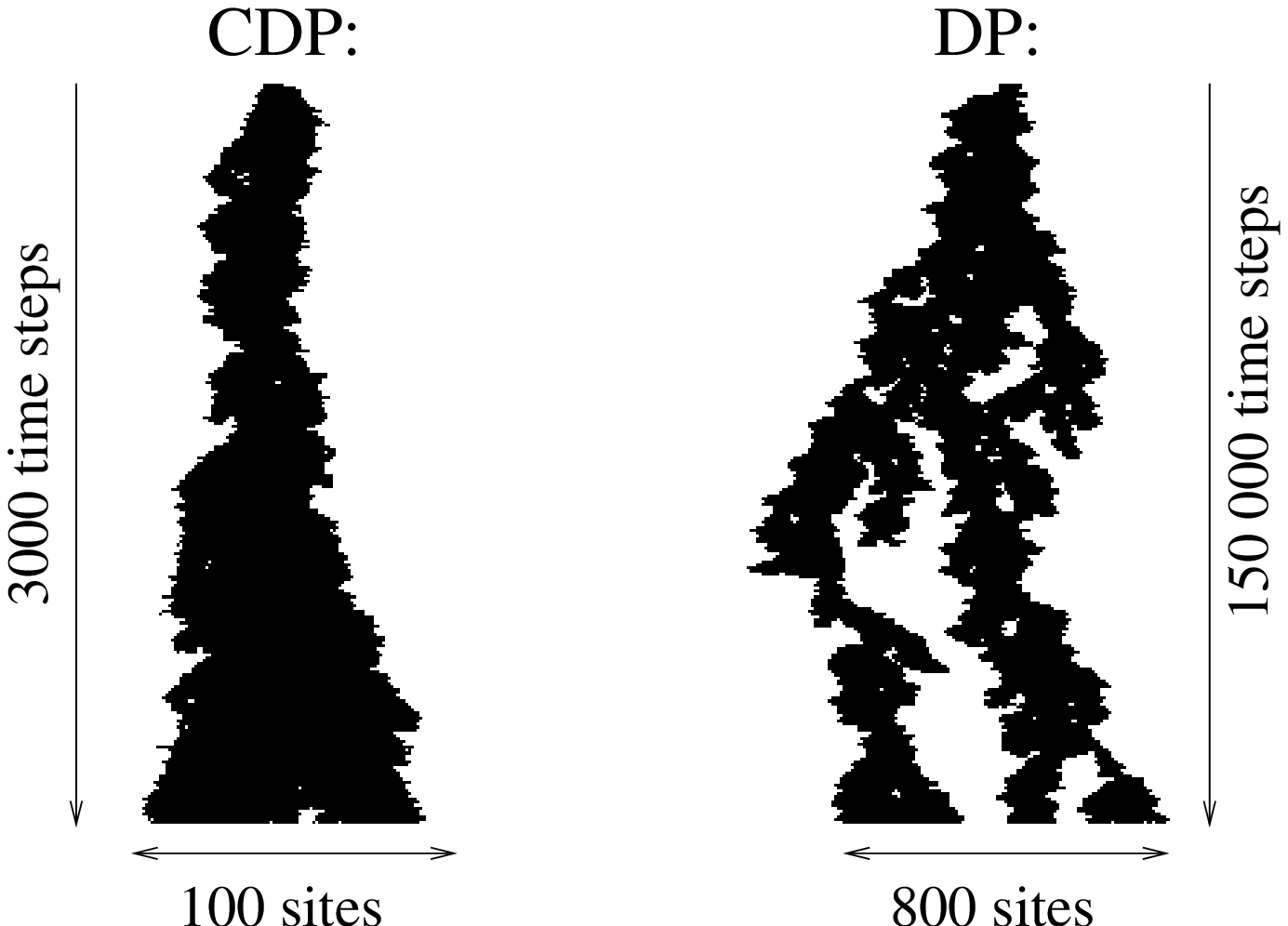}}
\caption{\label{FigDemoCrossover}}
\end{figure}

\begin{figure}
\epsfxsize=130mm
\centerline{\epsffile{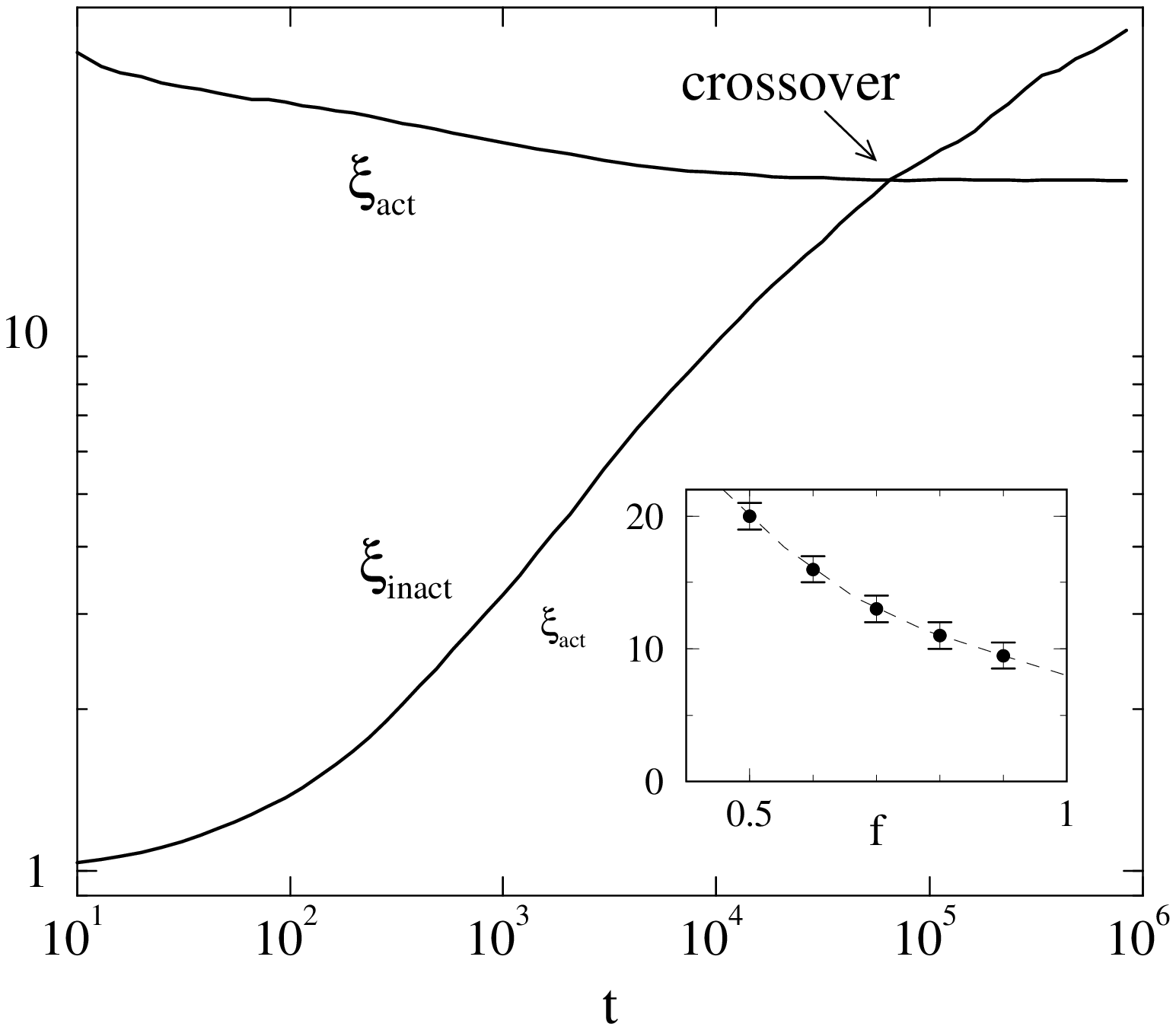}}
\caption{\label{FigIslands}}
\end{figure}


\begin{thebibliography}{99}

\bibitem{DP}
W.~Kinzel, in {\em Percolation Structures and Processes},
ed. G.~Deutscher, R.~Zallen, and J.~Adler,
Ann. Isr. Phys. Soc. {\bf 5}
(Adam Hilger, Bristol, 1983), p.~425.

\bibitem{Jensen}
Currently the most precise estimates for the DP exponents
are given in: I. Jensen, cond-mat/9906036 (unpublished).

\bibitem{Grassberger96}
P. Grassberger,
{\it Directed percolation: results and open problems},
preprint WUB 96-2 (1996), unpublished.


\bibitem{DouadyDaerr98}
S. Douady and A. Daerr, 
{\it Physics of Dry Granular Media}, H. J. Herrmann et al.,
eds., p. 339, Kluwer Academic Publishers, NY (1998). 

\bibitem{DaerrDouady99}
A. Daerr and S. Douady, Nature {\bf 399}, 241 (1999).

\bibitem{DomanyKinzel} E. Domany and W. Kinzel, 
Phys. Rev. Lett. {\bf 53}, 311 (1984).

\bibitem{TadicDhar97}
B. Tadic and D. Dhar, Phys. Rev. Lett. {\bf 79}, 1519 (1997).

\bibitem{Bak} 
P. Bak, C. Tang and K. Wiesenfeld,
Phys. Rev. Lett. {\bf 59}, 381 (1987).

\bibitem{DickmanTretyakov95}
R. Dickman and A. Yu. Tretyakov, Phys. Rev.
{\bf E 52}, 3218 (1995).

\bibitem{DPConjecture}
H. K. Janssen, Z. Phys. {\bf B 42}, 151 (1981);
P. Grassberger, Z. Phys. {\bf B 47}, 365 (1982).

\bibitem{GrassbergerTorre79}
P. Grassberger and A. de la Torre,
Ann. Phys. (NY) {\bf 122}, 373 (1979).

\bibitem{MDH96}
J. F. F. Mendes, R. Dickman, and H. Herrmann,
Phys. Rev. {\bf E 54}, R3071 (1996).

\bibitem{MoreiraDickman96}
A. G. Moreira and R. Dickman, Phys. Rev. {\bf E 54}, 1 (1996);
R. Dickman and A. G. Moreira, Phys. Rev. Lett. {\bf 57}, 1263 (1998).

\bibitem{Janssen97}
H. K. Janssen, Phys. Rev. {\bf E 55}, 6253 (1997).

\bibitem{CGM98}
R. Cafiero, A. Gabrielli, and M. A. Mu{\~n}oz, 
Phys. Rev. {\bf E 57}, 5060 (1998).

\bibitem{WACH98}
I. Webman, D. ben-Avraham, A. Cohen, and S. Havlin,
Phil. Mag. {\bf B 77}, 1401 (1998).

\bibitem{Jensen96}
I. Jensen, Phys. Rev. Lett. {\bf 77}, 4988 (1996).

\bibitem{Janssen81}
H. K. Janssen, Z. Phys. {\bf B 42},  151 (1981).

\bibitem{BouchaudCates98}
J. P. Bouchaud and M. E. Cates, Granular Matter {\bf 1}, 101 (1998).

\end{thebibliography}
\end{document}